\documentclass[11pt, a4paper]{article}
\usepackage[utf8]{inputenc}
\usepackage{amsmath,amssymb,amsthm}
\usepackage{graphicx}
\usepackage{geometry}
\usepackage{enumitem}
\usepackage{hyperref}
\usepackage{dblfloatfix}
\usepackage{latexsym}
\usepackage{booktabs}
\usepackage{color}
\usepackage{todonotes}
\usepackage{dsfont}
\usepackage{subcaption}
\usepackage{float}

\title{A New Perspective on Drawing\\ Venn Diagrams for Data Visualization}
\author{Bálint Csanády\\\\Institute of Mathematics, ELTE Eötvös Loránd University, Budapest, Hungary}
\date{}

\begin{document}
\maketitle

\begin{abstract}
We introduce \emph{VennFan}, a method for generating $n$-set Venn diagrams based on the polar coordinate projection of trigonometric boundaries, resulting in Venn diagrams that resemble a set of fan blades.
Unlike most classical constructions, our method emphasizes readability and customizability by using shaped sinusoids and amplitude scaling.
We describe both sine- and cosine-based variants of \emph{VennFan}
and propose an automatic label placement heuristic tailored to these fan-like layouts.
\emph{VennFan} is available as a Python package\footnote{https://pypi.org/project/vennfan/}. 
\end{abstract}

\section{Introduction}

\begin{figure}[htbp]
     \centering
     \begin{subfigure}[b]{.22\textwidth}
        \centering
        \includegraphics[width=\textwidth]{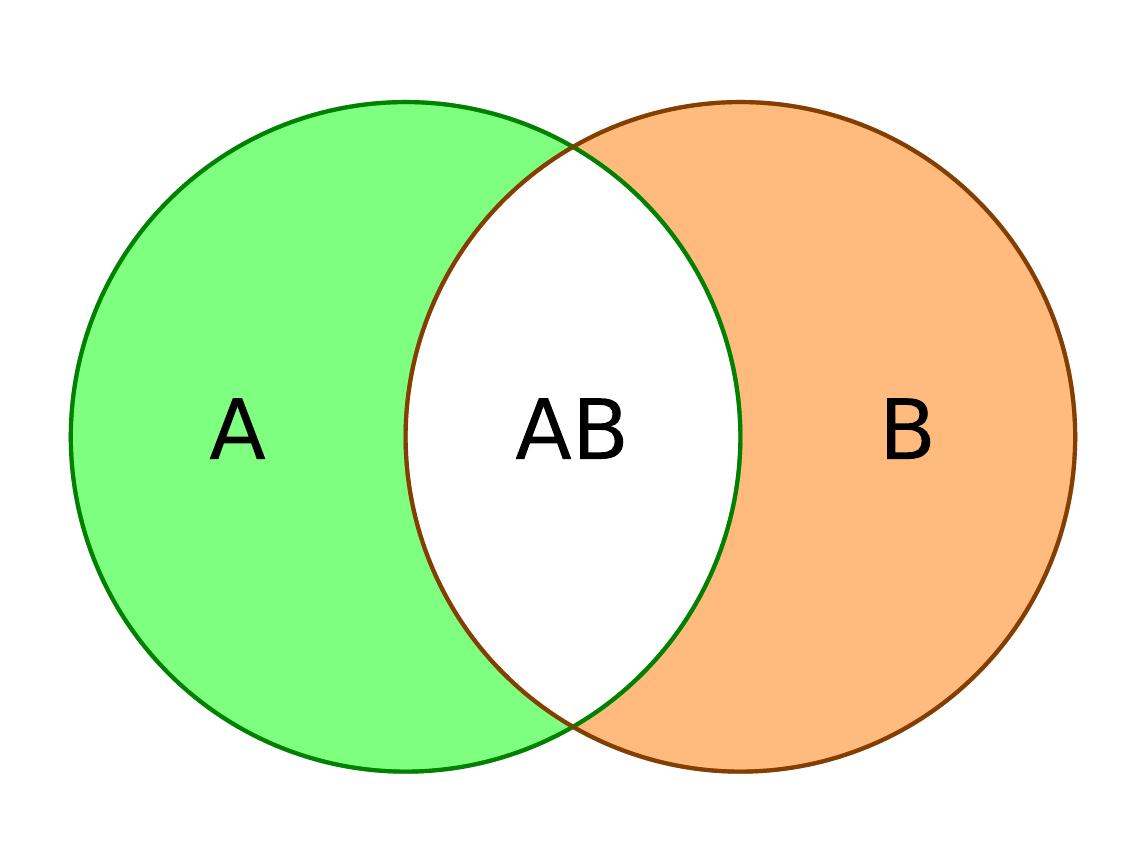}
        \subcaption{$n=2$.}
     \end{subfigure}
     \hfill
     \begin{subfigure}[b]{.22\textwidth}
        \centering
        \includegraphics[width=\textwidth]{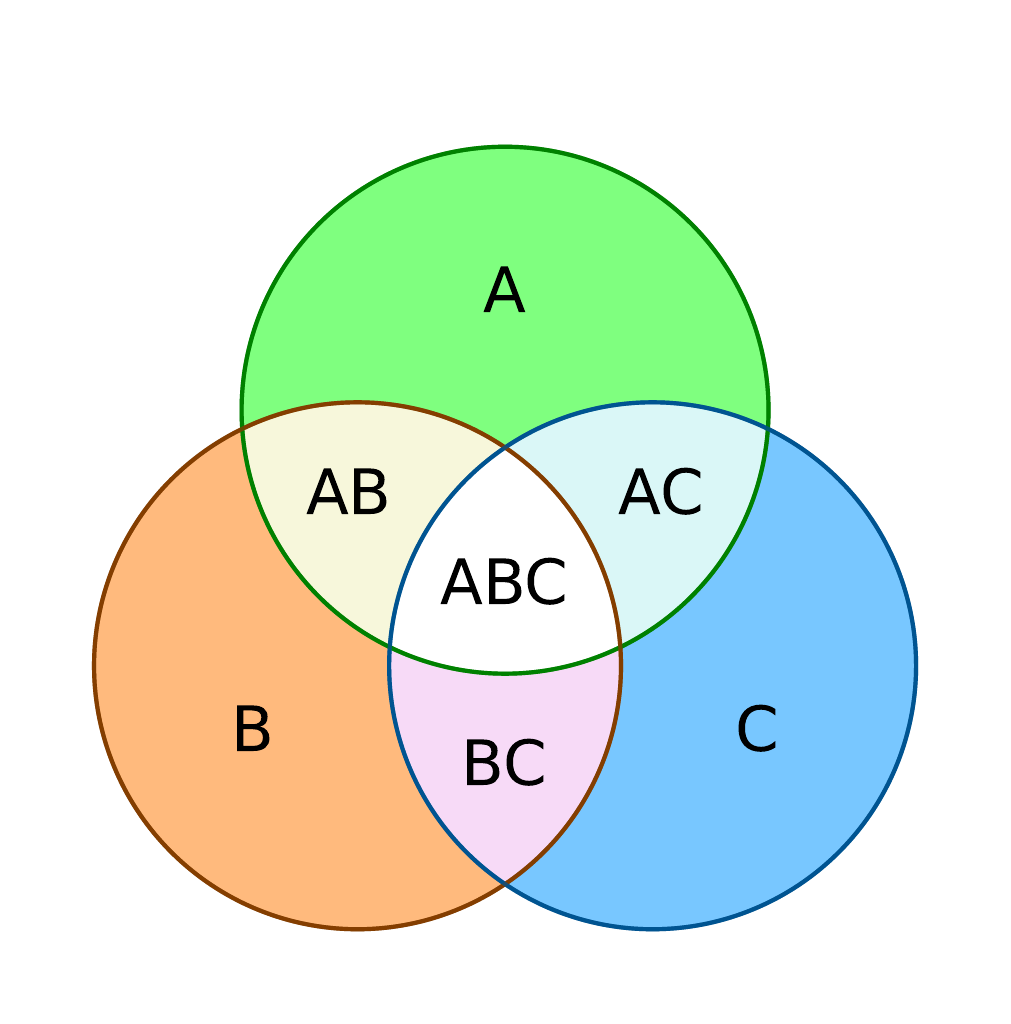}
        \subcaption{$n=3$.}
     \end{subfigure}
     \hfill
     \begin{subfigure}[b]{.22\textwidth}
        \centering
        \includegraphics[width=\textwidth]{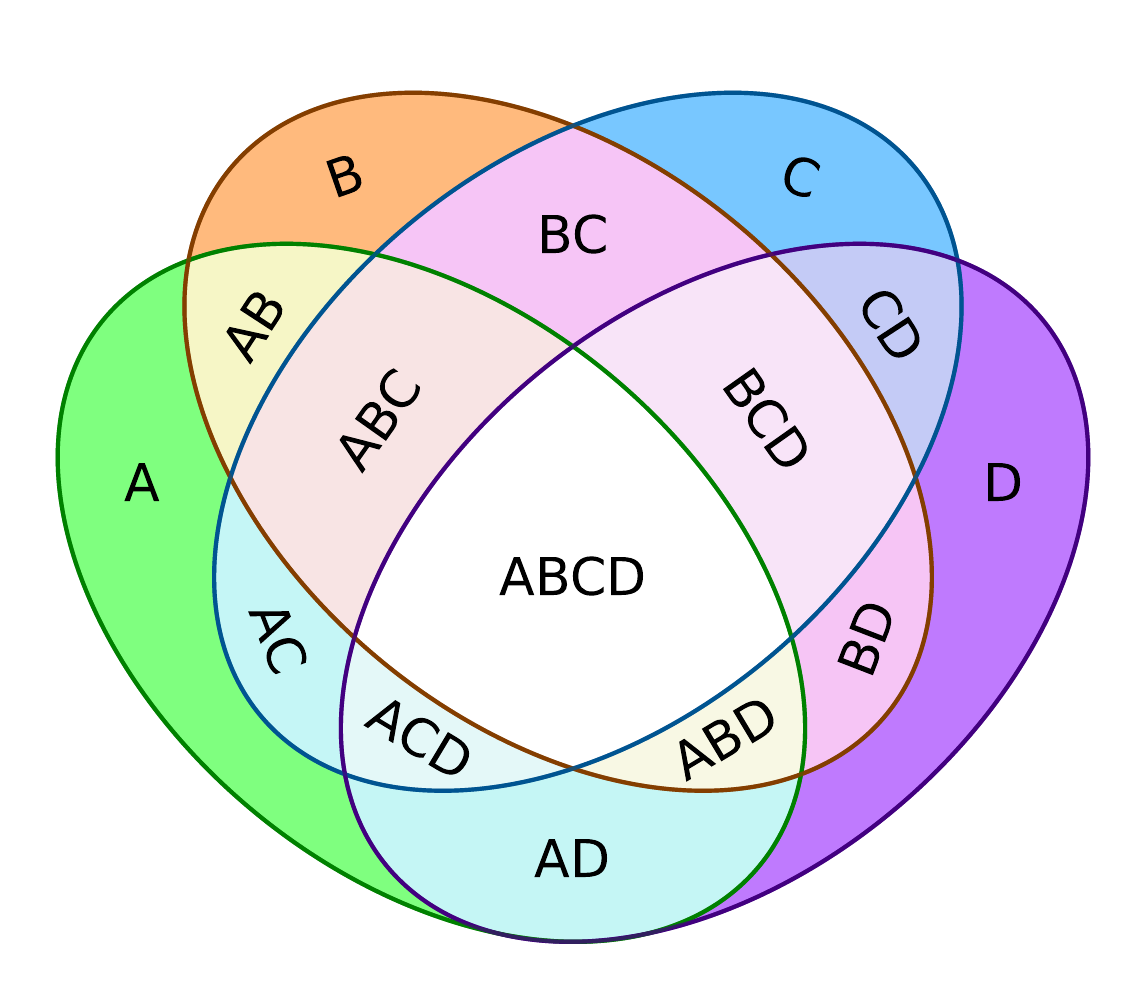}
        \subcaption{$n=4$.}
     \end{subfigure}
     \hfill
     \begin{subfigure}[b]{.22\textwidth}
        \centering
        \includegraphics[width=\textwidth]{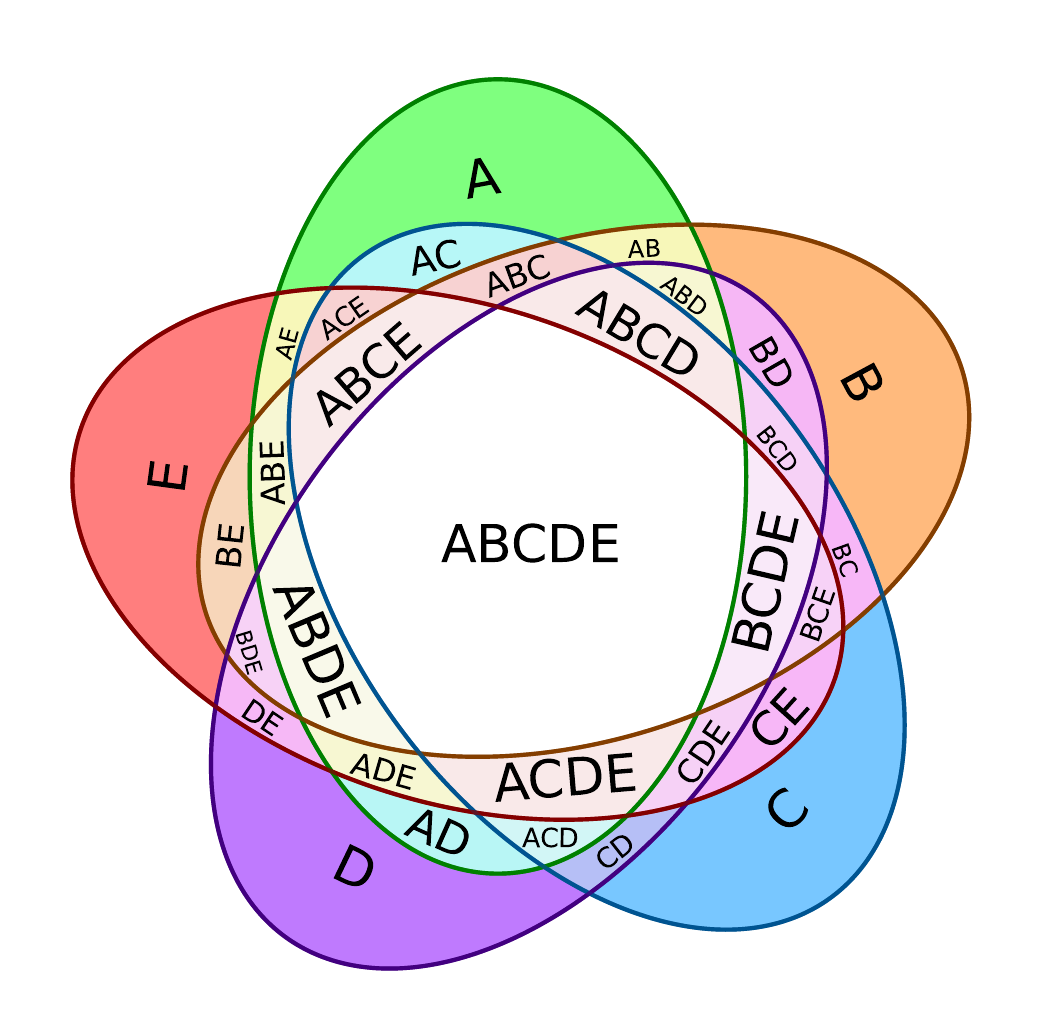}
        \subcaption{$n=5$.}
     \end{subfigure}
     \caption{Ellipsoidal Venn diagrams for $2 \le n \le 5$.}
    \label{fig:ellipsoidal}
\end{figure}

Venn diagrams visualize all possible logical relations between $n$ sets.
For $n<6$, they can be realized by overlapping ellipses (Figure \ref{fig:ellipsoidal}) \cite{Venn1880,gruenbaumIndependent}.
Using Euler’s formula together with the fact that two ellipses can meet in at most four points, Grünbaum showed that no Venn diagram with six or more sets can be drawn using ellipses \cite{grunbaum1992venn}.
Therefore, due to the exponential nature of Venn diagrams, the readability of known constructions deteriorates rapidly as $n$ increases beyond $5$.

The challenge is not only to ensure topological completeness (that all $2^n$ intersections are represented), but also to support the presentation of data in a form that remains interpretable.
Ultimately, what counts as “readable’’ is subjective and up to personal preference.
Consequently, several additional constraints on Venn diagrams have been proposed with the implicit goal of improving readability.
These constraints lead to interesting combinatorial and geometric problems in their own right, and deserve attention independently of their effectiveness for data visualization.

\begin{figure}[htbp]
     \centering
     \begin{subfigure}[b]{.40\textwidth}
        \centering
        \includegraphics[width=\textwidth]{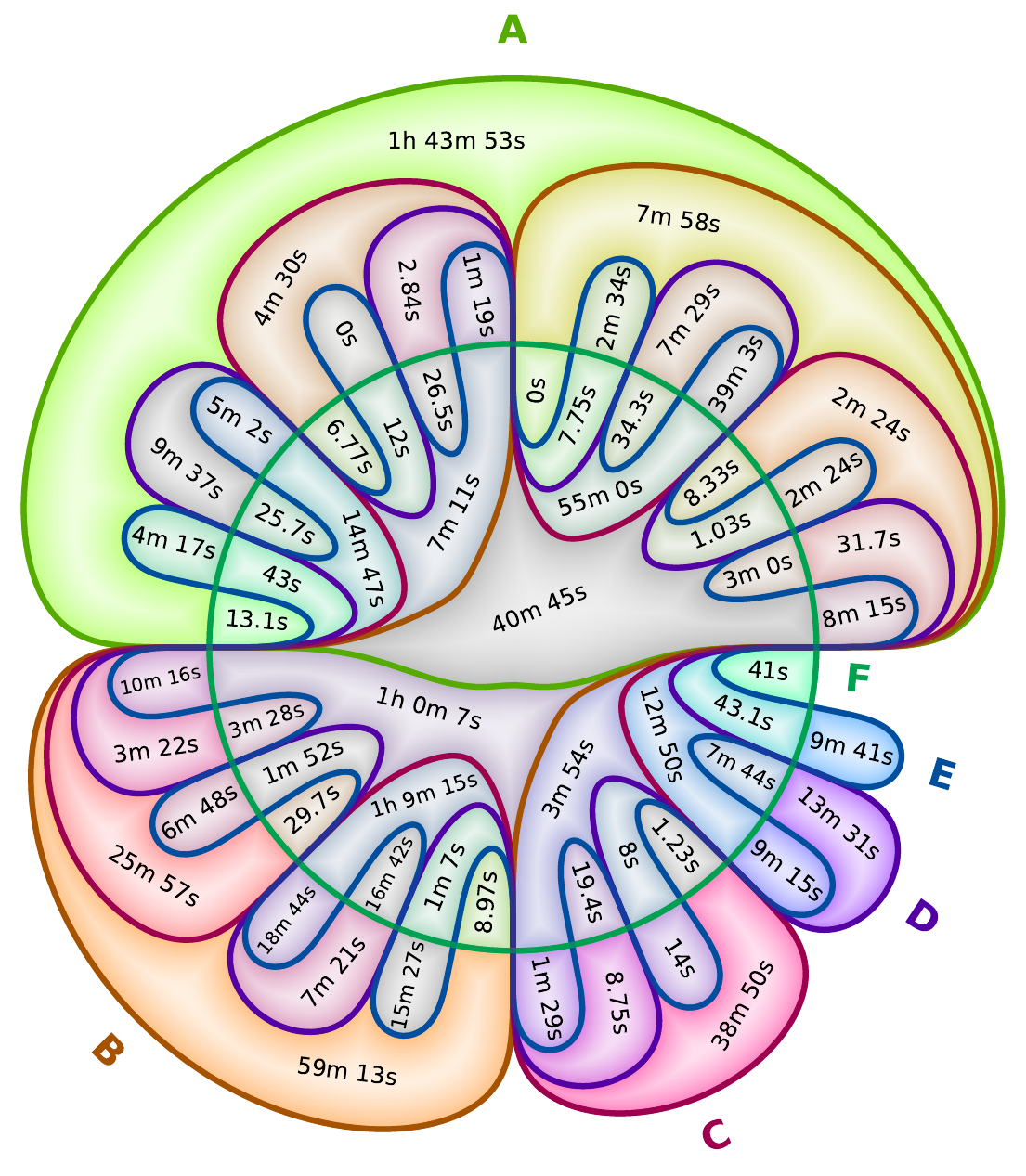}
        \subcaption{Sine-based.}
     \end{subfigure}
     \hfill
     \begin{subfigure}[b]{.46\textwidth}
        \centering
        \includegraphics[width=\textwidth]{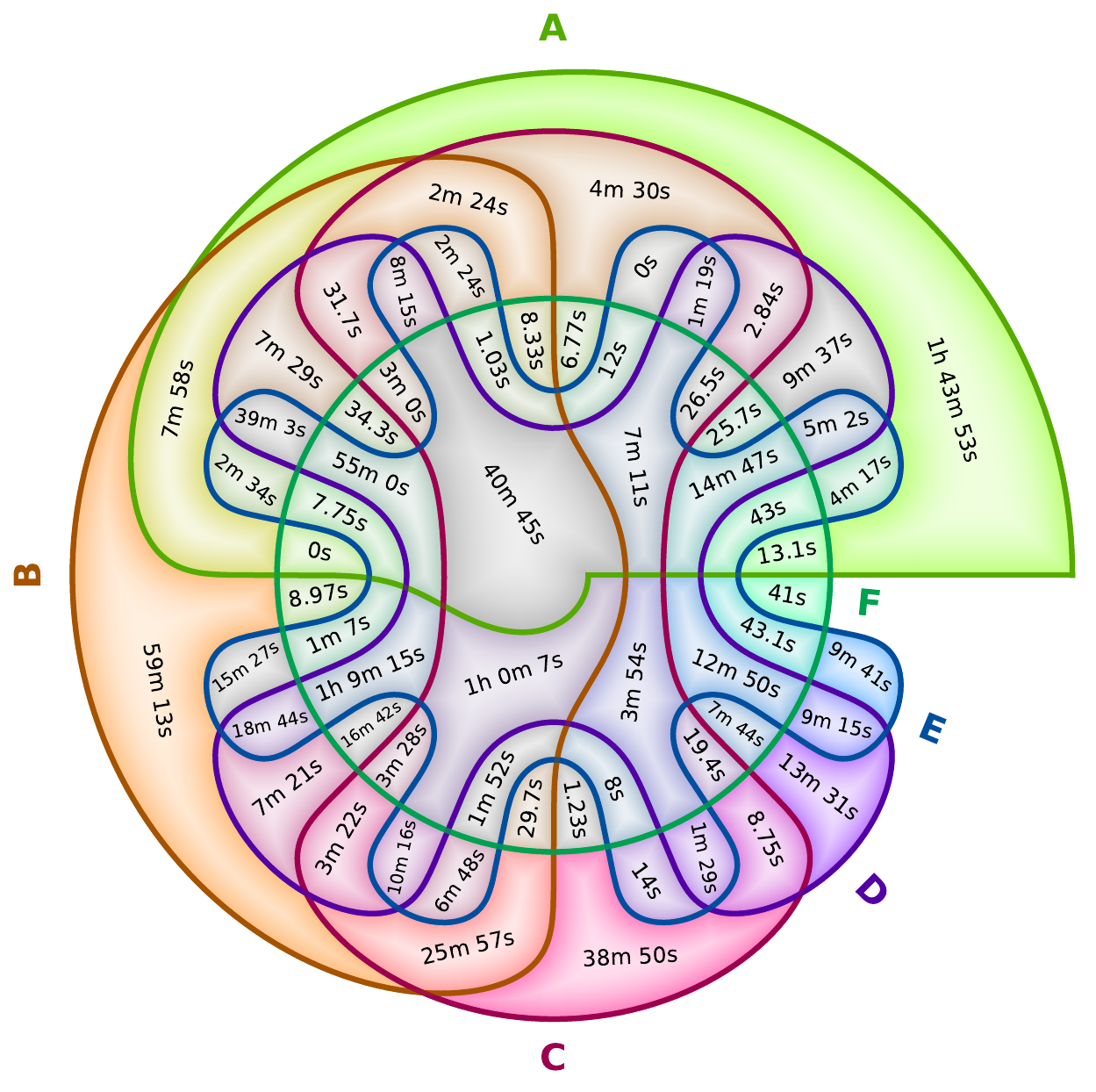}
        \subcaption{Cosine-based.}
     \end{subfigure}
     \caption{\emph{VennFan} diagrams of overlaps between six different seizure annotators in EEG data.
     }
    \label{fig:vennfan}
\end{figure}

In this work however, we argue that relaxing most of these constraints allows the construction of Venn diagrams that are, in many situations, better suited for presenting data.
One domain where such data naturally arises, and where Venn diagrams may be particularly useful, is biology and medicine.
For time series data annotated by multiple experts, one may be interested in the overlap structure between the different sets of annotations.
For example, annotations may come from several human experts (e.g., clinicians) or from automated detection software that mark abnormalities within the data, such as epileptic seizures in EEG recordings.
Our construction, \emph{VennFan}, provides a flexible method for representing such information with clearly readable labels and visually distinct classes (Figure~\ref{fig:vennfan}).

\section{Background}
Following Grünbaum and Ruskey~\cite{gruenbaumIndependent,ruskey2005survey}, let $C = \{C_0,\dots,C_{n-1}\}$ be a family of simple closed curves in the plane, each representing one of $n$ sets.
For each $i$, let $X_i$ denote either the bounded interior or the unbounded exterior of $C_i$.
We call $C$ an \emph{independent family} if every one of the $2^n$ intersections
\[
  X_0 \cap X_1 \cap \cdots \cap X_{n-1}
\]
is non-empty.
Each connected component of such an intersection is called a \emph{region}.
If every one of the $2^n$ intersections consists of a single region, we call the independent family a \emph{Venn diagram}.
A Venn diagram is called \emph{simple} if at most two boundaries meet at any point \cite{carrollkgons}.

\begin{figure}[htbp]
     \centering
     \begin{subfigure}[b]{.35\textwidth}
        \centering
        \includegraphics[width=\textwidth]{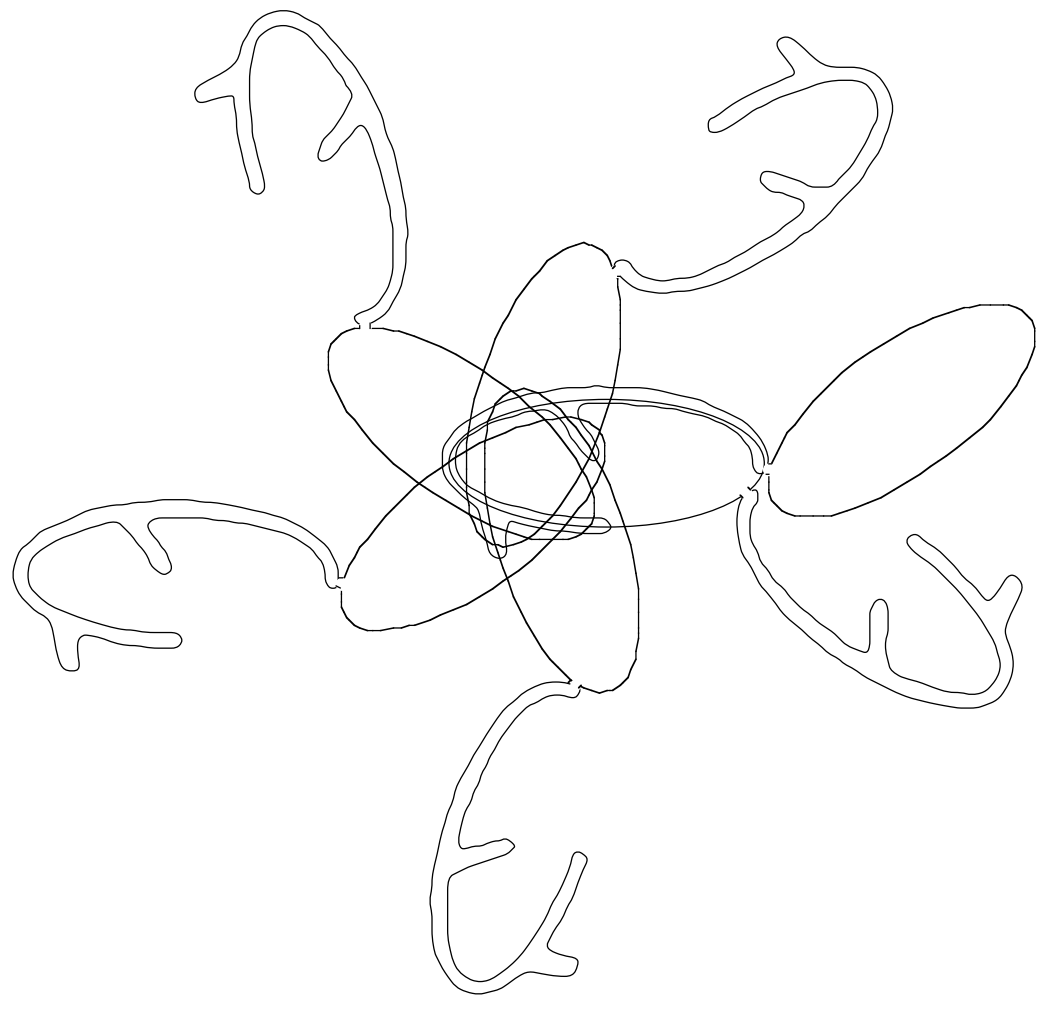}
        \subcaption{Grünbaum's nice diagram.}
        \label{fig:Nice}
     \end{subfigure}
     \hfill
     \begin{subfigure}[b]{.4\textwidth}
        \centering
        \includegraphics[width=\textwidth]{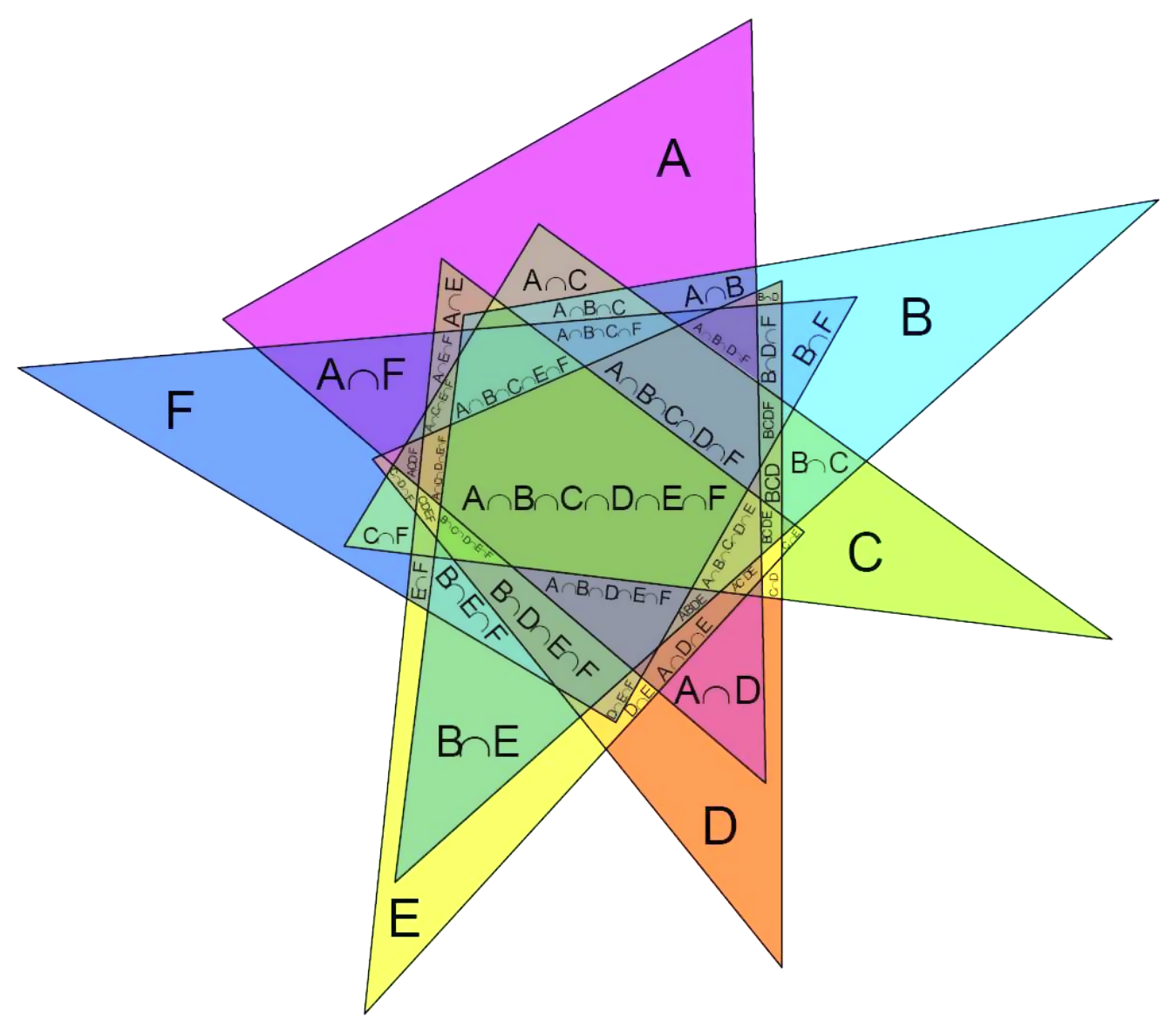}
        \subcaption{Carroll's simple triangle diagram.}
        \label{fig:Carroll}
     \end{subfigure}
     \caption{Nice and simple Venn diagrams for $n=6$.}
\end{figure}

A natural requirement motivated by the $n \le 5$ ellipsoidal constructions would be to represent all of the curves by congruent shapes; such diagrams are called \emph{nice} \cite{grunbaum1992venn2} or \emph{congruent} \cite{ruskey2005survey} (Figure ~\ref{fig:Nice}).
Another natural way to generalize is to keep the shapes as elementary as possible: for $n=6$ it is feasible to draw a simple Venn diagram using triangles. Figure~\ref{fig:Carroll} shows Carroll's construction found by computer search~\cite{carroll2000drawing}.
The ultimate marriage for these two ideas --- a Venn diagram on six sets
drawn with congruent equilateral triangles --- appears to remain open~\cite{ruskey2005survey, gruenbaumIndependent}.
Although these conditions are easily motivated by looking at the familiar diagrams for $n \le 5$, their success in the quest for readable data visualizations is debatable.

\begin{figure}[htbp]
     \centering
     \begin{subfigure}[b]{.48\textwidth}
        \centering
        \includegraphics[width=\textwidth]{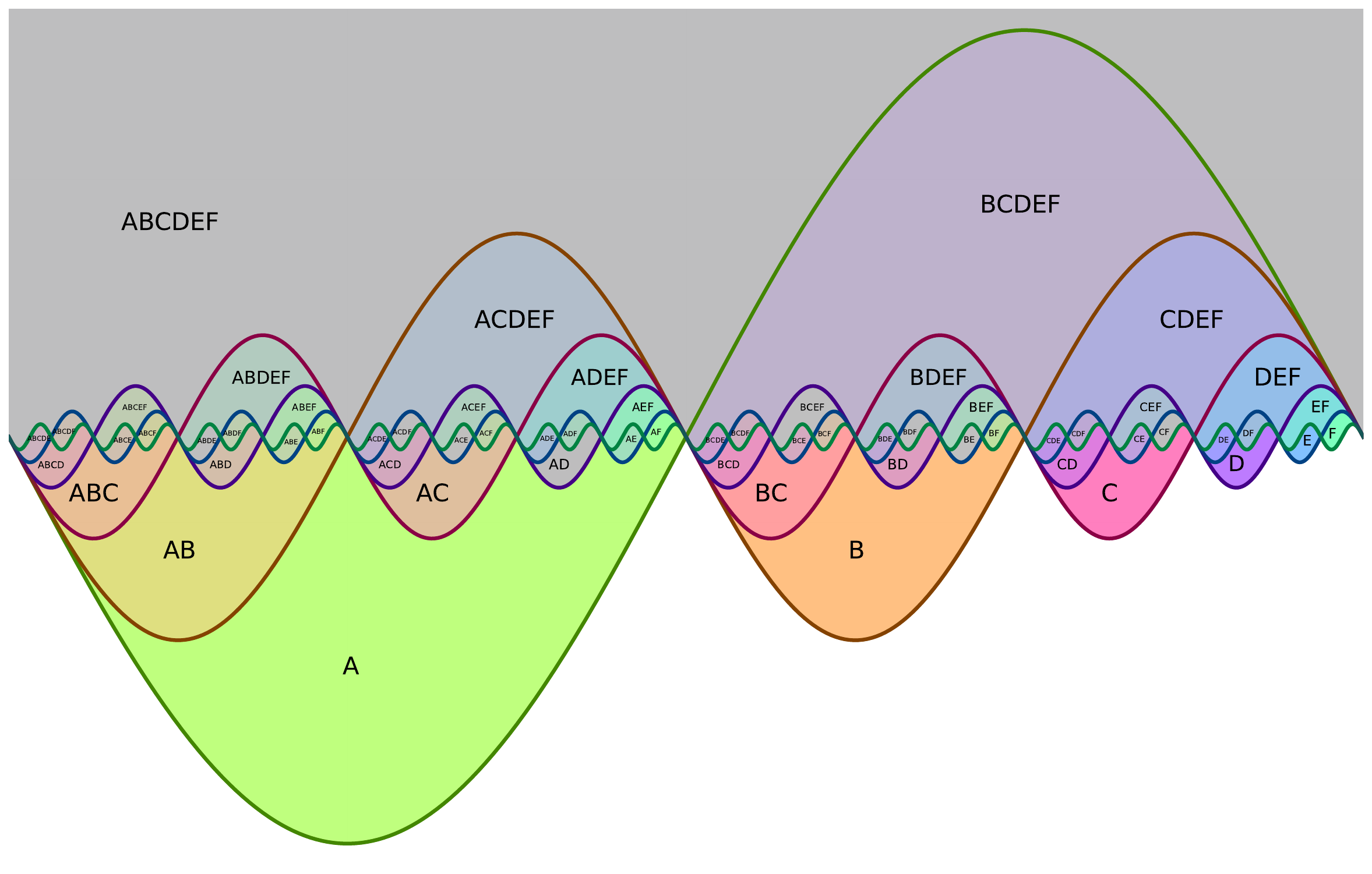}
        \subcaption{Smith's diagram.}
        \label{fig:Smith}
     \end{subfigure}
     \hfill
     \begin{subfigure}[b]{.42\textwidth}
        \centering
        \includegraphics[width=\textwidth]{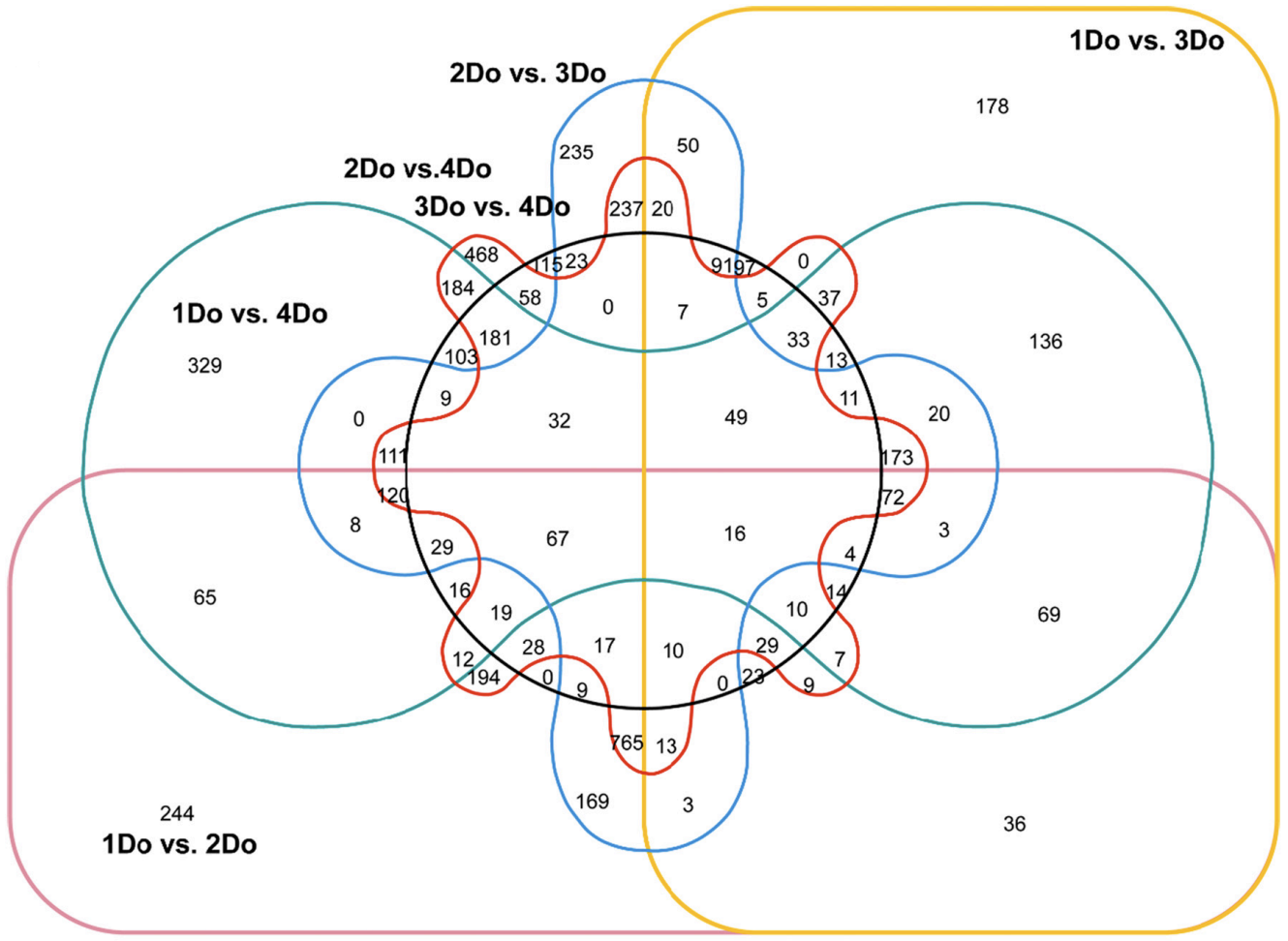}
        \subcaption{Edwards' cogwheel diagram.}
        \label{fig:Edwards}
     \end{subfigure}
     \caption{Examples of trigonometric (Smith) and cogwheel (Edwards) diagrams.}
\end{figure}

An elegant general Venn-like construction for arbitrary $n$ uses trigonometric curves; it goes back to ideas of C.~A.~B.~Smith and was later discussed by Edwards~\cite{edwards2004cogwheels,ruskey2005survey}.
In this classical construction,
\begin{align}
  f_i(x) = \sin(2^i x)/2^i,
  \quad x \in [-\pi,\pi],
  \quad i = 0,1,\ldots,n-1,
\end{align}
the amplitude decays exponentially with $i$, and when it is drawn as a graph in the plane the boundaries are not closed curves, so they do not form a planar Venn diagram in the usual sense (Figure~\ref{fig:Smith}).
Both of these issues are addressed by \emph{VennFan}, which is inspired by Smith's construction.

Finally, a particularly influential way of drawing Venn diagrams for $n > 5$ is the spherical construction pioneered by Edwards~\cite{edwards1989venn,edwards2004cogwheels}, which yields his well-known \emph{cogwheel} diagrams after projection to the plane.
His design, although arguably the state-of-the-art option for representing data, also suffers from a rapid decrease of region sizes as $n$ grows, and its planar appearance can look somewhat arbitrary at first glance.
Figure \ref{fig:Edwards} shows an example of cogwheel diagrams being used to represent gene expressions by Yuan et al. \cite{yuan2022transcriptome}.
As we will see, the cosine-based version of \emph{VennFan} is topologically equivalent to Edwards' construction, and thus offers an alternative, more tunable way to realize the same combinatorial structure.

\section{The \emph{VennFan} Construction}

\begin{figure}[htbp]
     \centering
     \begin{subfigure}[b]{.48\textwidth}
        \centering
        \includegraphics[width=\textwidth]{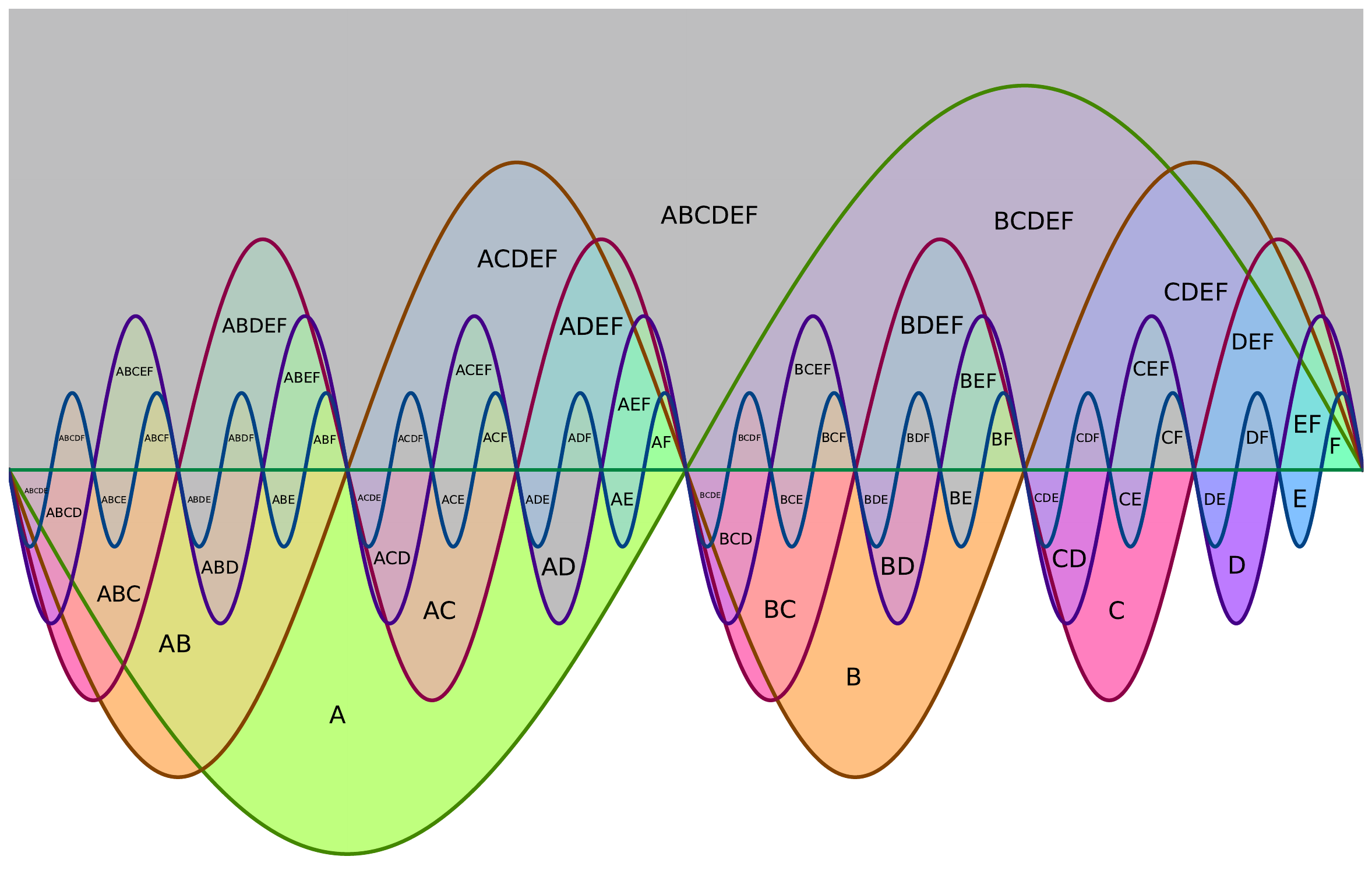}
        \subcaption{$p=1$.}
        \label{fig:nop_sine_linear_N6}
     \end{subfigure}
     \hfill
     \begin{subfigure}[b]{.48\textwidth}
        \centering
        \includegraphics[width=\textwidth]{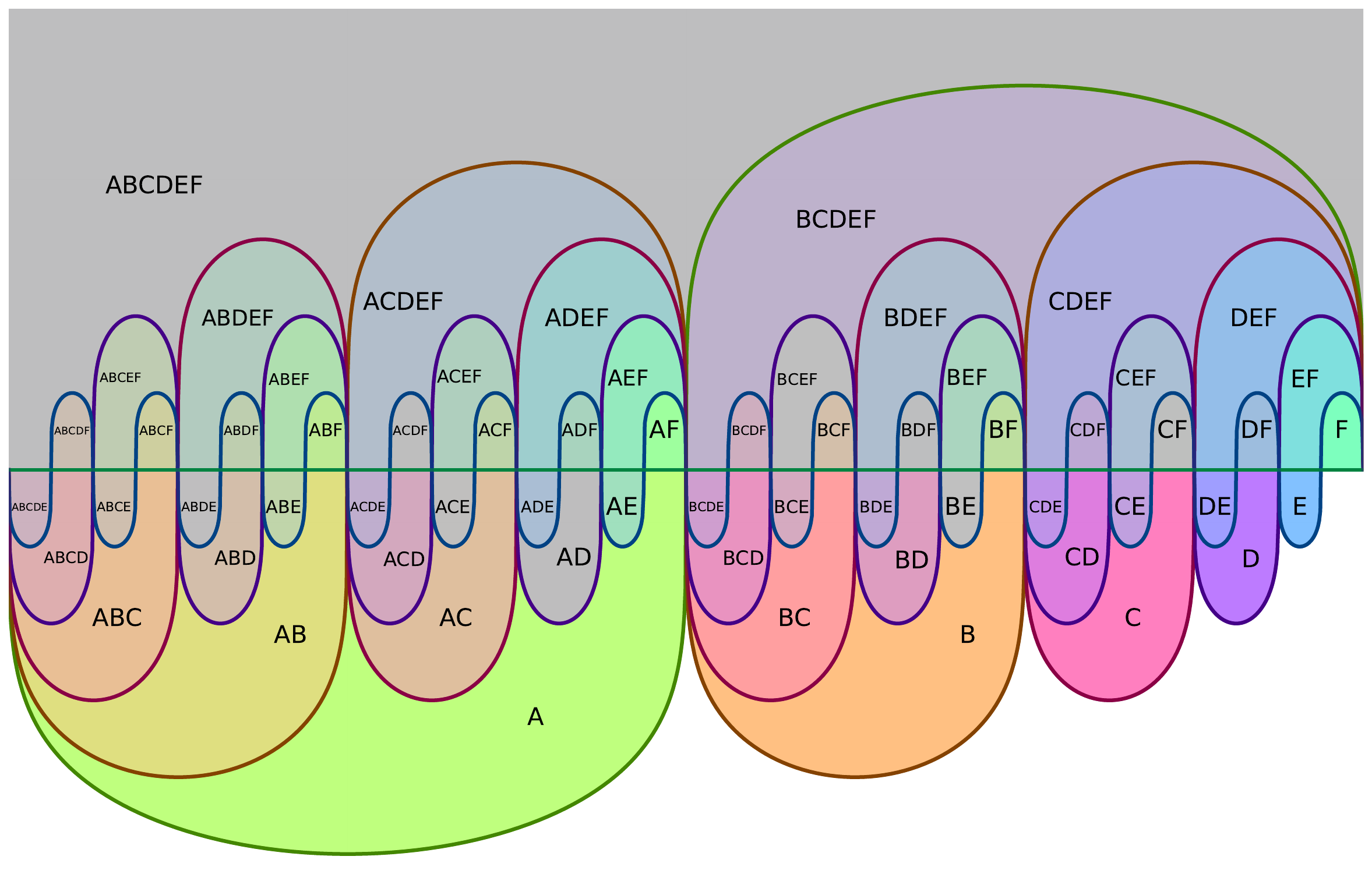}
        \subcaption{$p=1/5$.}
        \label{fig:sine_linear_N6}
     \end{subfigure}
     \caption{Linear decay on the sinusoidal Smith diagram.}
\end{figure}

The first step is to overcome the exponential amplitude decay present in Smith's construction.
For example, instead of exponential amplitude decay, we can introduce linear decay $\lambda(i)=\tfrac{n-1-i}{n}$.
However, if we do this, the curves will collide (Figure \ref{fig:nop_sine_linear_N6}).
To mitigate these collisions, we \emph{shape} the curves by raising them to a power with exponent $p<1$, leading to Equation~\ref{eq:sin_linear}:

\begin{align}
f_i(x) = \lambda(i)\,\operatorname{sgn}\left(\sin(2^{i} x)\right)\,\left|\sin(2^{i} x)\right|^{p},
\quad x &\in [-\pi,\pi],
\quad i = 0,\ldots,n-1.
\label{eq:sin_linear}
\end{align}

Note, that as we decrease $p<1$ the shape of the sine waves become ``fatter'' (Figure~\ref{fig:sine_linear_N6}), with $p\rightarrow 0$ tending to a square wave.

\begin{figure}[H]
     \centering
     \begin{subfigure}[b]{.465\textwidth}
        \centering
        \includegraphics[width=\textwidth]{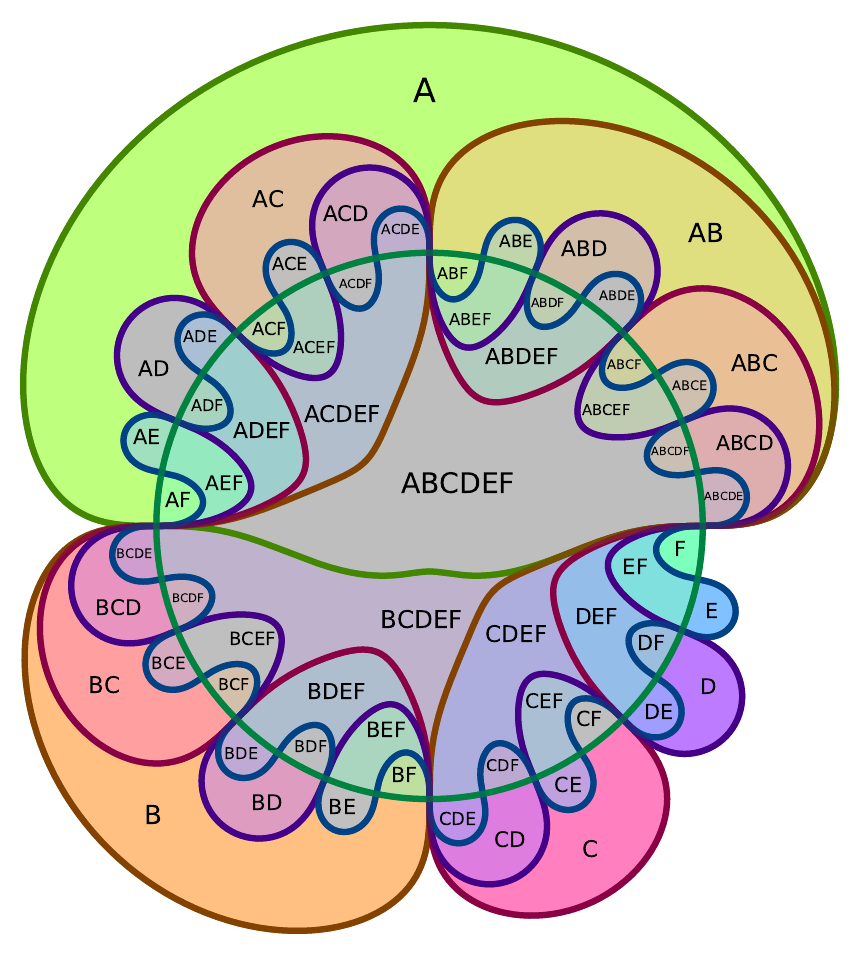}
        \subcaption{Linear decay, $p=1/3$.}
        \label{fig:vennfan_sine_linear_N6}
     \end{subfigure}
     \hfill
     \begin{subfigure}[b]{.465\textwidth}
        \centering
        \includegraphics[width=\textwidth]{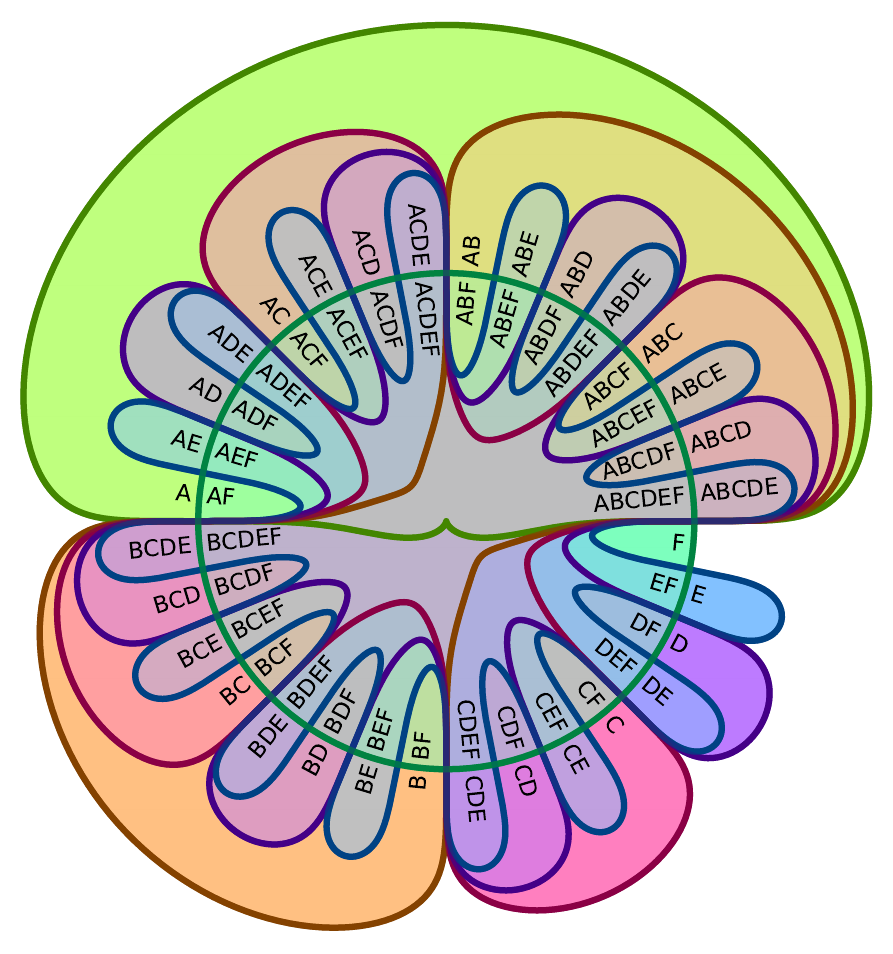}
        \subcaption{$b=4/5,\,p=1/5$.}
        \label{fig:vennfan_sine_N6}
     \end{subfigure}
     \caption{Sinusoidal \emph{VennFan} diagrams for $n=6$ with linear and modified exponential decay.}
\end{figure}

The second step is to turn the unbounded outer region into a bounded one and to convert the boundaries into simple closed curves.
We achieve this by projecting the rectangle $[-\pi,\pi]\times[-1,1]$ onto the unit disc by treating $x$ as the polar angle $\vartheta$ and $f_i(x)$ as a radial deviation from radius $1$ (Figure~\ref{fig:vennfan_sine_linear_N6}).
More precisely, we draw the parametric curve:
\begin{align}
  g_i(\vartheta) &= \big(\left(1 + f_i(\vartheta)\right)\cos \vartheta,\; \left(1 + f_i(\vartheta)\right)\sin \vartheta\big),
  \quad \vartheta \in [-\pi,\pi],\quad i=0,\ldots,n-1.
  \label{eq:polar_projection}
\end{align}
Notice that $\left|f_i(\vartheta)\right| < 1$, which prevents $g_i(\vartheta)$ from passing through the origin.

Instead of linear decay we can bring back the exponential decay, but with a base $1/2<b<1$:
\begin{align}
\lambda(i)&=b^{i+\varepsilon},\quad i = 0,1,\ldots,n-2,\\
\lambda(n-1)&=0.
\label{eq:sin_exp}
\end{align}
If the shape parameter $p$ is small enough, then as we increase $b$, we create more space for text labels around the circle (Figure~\ref{fig:vennfan_sine_N6}).
Here $\lambda(n-1)=0$ keeps the last boundary the $y=0$ line, which is mapped to the unit circle by the projection.
Additionally, $\varepsilon > 0$ ensures that $f_i(x)<1$, and we do not cross the origin when projecting.
In practice, we can use $\varepsilon=0$, or even $\varepsilon \gg 0$ to adjust the appearance of the inside of the projected diagram.

An arguably more visually pleasing alternative to the modified exponential amplitude decay is to modify the linear decay as:
\begin{align}
\lambda_{\delta,\epsilon}(i)&=\tfrac{\delta+\varepsilon-1}{n-2}i-\varepsilon+1,\quad i = 0,1,\ldots,n-2,\\
\lambda_{\delta,\epsilon}(n-1)&=0.
\label{eq:sin_linear_mod}
\end{align}
Here $\delta$ controls the size of the smallest ``fan blades'', and $\varepsilon$ the size of the first step we take in the amplitude decay process.
In case of the basic linear decay ($\lambda(i)=\tfrac{n-1-i}{n}$), these parameters were the same: $\delta=\varepsilon=1/n$.

\begin{figure}[htbp]
     \centering
     \begin{subfigure}[b]{.56\textwidth}
        \centering
        \includegraphics[width=\textwidth]{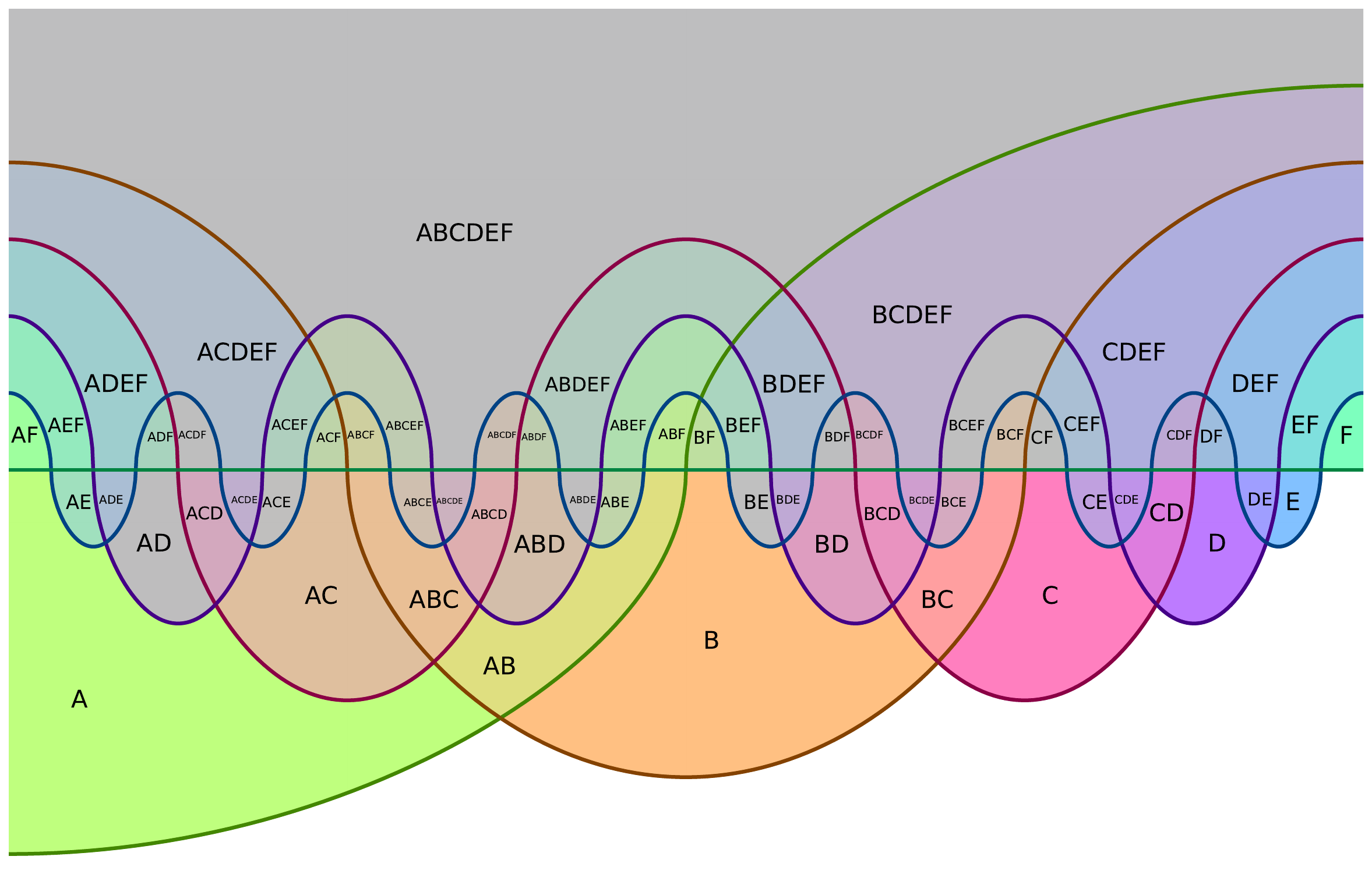}
        \subcaption{}
        \label{fig:vennfan_cosine}
     \end{subfigure}
     \hfill
     \begin{subfigure}[b]{.40\textwidth}
        \centering
        \includegraphics[width=\textwidth]{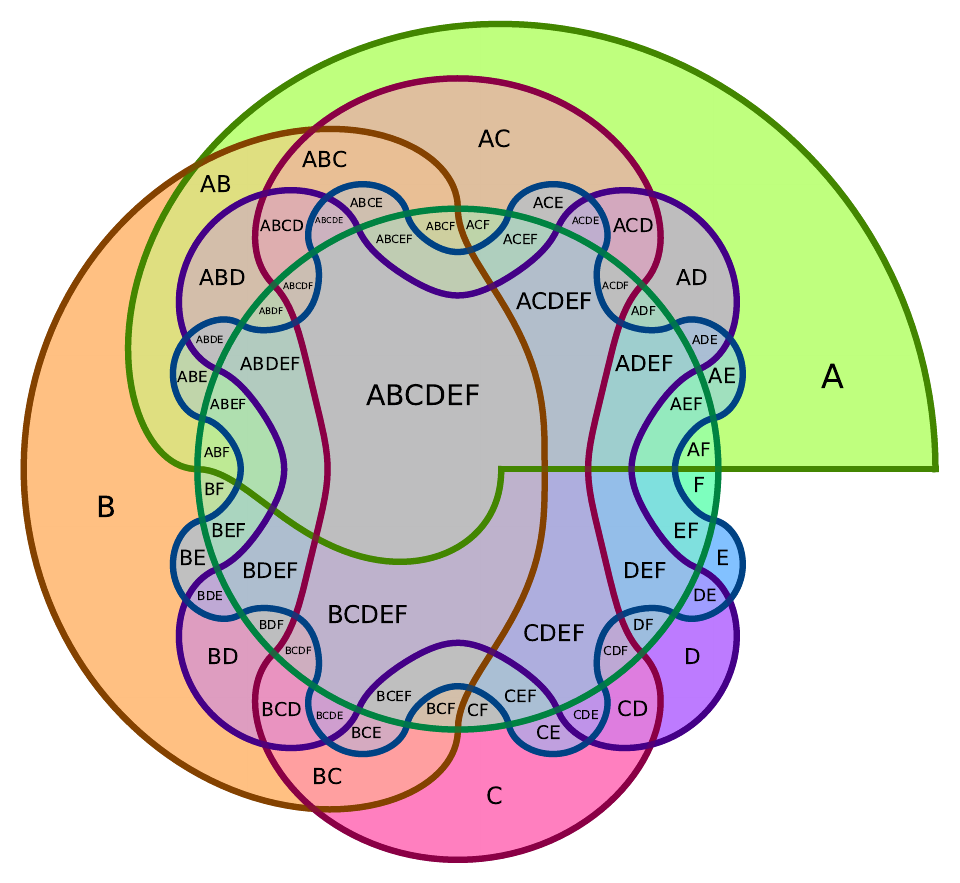}
        \subcaption{}
        \label{fig:vennfan_cosine_proj}
     \end{subfigure}
     \caption{Cosine-based Venn-like diagram with linear scaling (a) and its projection (b) with $p=1/2$.}
\end{figure}

Instead of sine functions, we can also use cosines to form a Venn-like diagram \cite{weston2009symmetries}.
Here we can also apply alternative scaling after shaping the curves (Equation~\ref{eq:cos_linear}, Figure~\ref{fig:vennfan_cosine}).
We use $x \in [2\pi,4\pi]$ (as opposed to $x \in [0,2\pi]$) to preserve the left-to-right order in which the sets are drawn.
The resulting \emph{VennFan} diagram turns out to be a planar rendering of Edwards’ cogwheel construction;
the resemblance is especially clear with linear scaling (Figure~\ref{fig:vennfan_cosine_proj}).
\begin{align}
f_i(x) = \lambda(i)\,\operatorname{sgn}\left(\cos(2^{i-1} x)\right)\,\left|\cos(2^{i-1} x)\right|^{p},
\quad x \in [2\pi,4\pi],
\quad i = 0,1,\ldots,n-1.
\label{eq:cos_linear}
\end{align}

\subsection{Empirical Properties}
\begin{figure}[htbp]
    \centering
     \begin{subfigure}[b]{.45\textwidth}
        \centering
        \includegraphics[width=\textwidth]{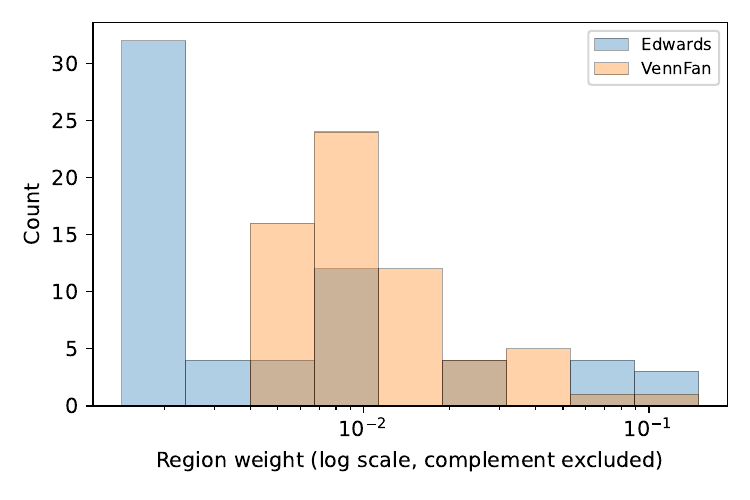}
        \subcaption{$n=6$.}
     \end{subfigure}
     \hfill
     \begin{subfigure}[b]{.45\textwidth}
        \centering
        \includegraphics[width=\textwidth]{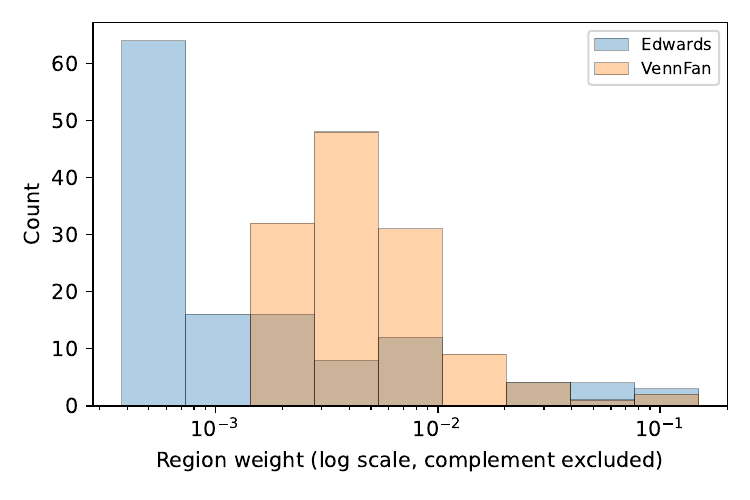}
        \subcaption{$n=7$.}
     \end{subfigure}
     \caption{Histograms of normalized region areas (weights) for Edwards’ cogwheel diagram and cosine-based \emph{VennFan} (modified linear amplitude decay $p=1/5,\,\delta=1/3,\,\varepsilon=1/9$).}
     \label{fig:area_hist}
\end{figure}

A major limitation of classical constructions for large $n$ is that the areas of the regions shrink rapidly, which makes it difficult to place readable labels.
To quantify this effect, in Figure~\ref{fig:area_hist} we compare the normalized region-area distributions of Edwards’ cogwheel diagram and the cosine-based \emph{VennFan}.
Although the two diagrams are topologically equivalent, the \emph{VennFan} construction produces a noticeably less extreme spread of region sizes, making it a more data-friendly planar realization of the same combinatorial structure.
To obtain accurate region areas for Edwards’ diagram we reimplemented his original spherical construction and its stereographic projection (Appendix~\ref{appx:Edwards}).
In principle, by pushing the parameters of \emph{VennFan} to extreme values (e.g.\ $p$ close to zero and $\lambda$ close to one), the region areas can be made nearly equal.
For a more realistic comparison, however, we report results for moderate parameter choices that preserve the overall look of the diagrams.

\begin{figure}[H]
     \centering
     \begin{subfigure}[b]{.45\textwidth}
        \centering
        \includegraphics[width=\textwidth]{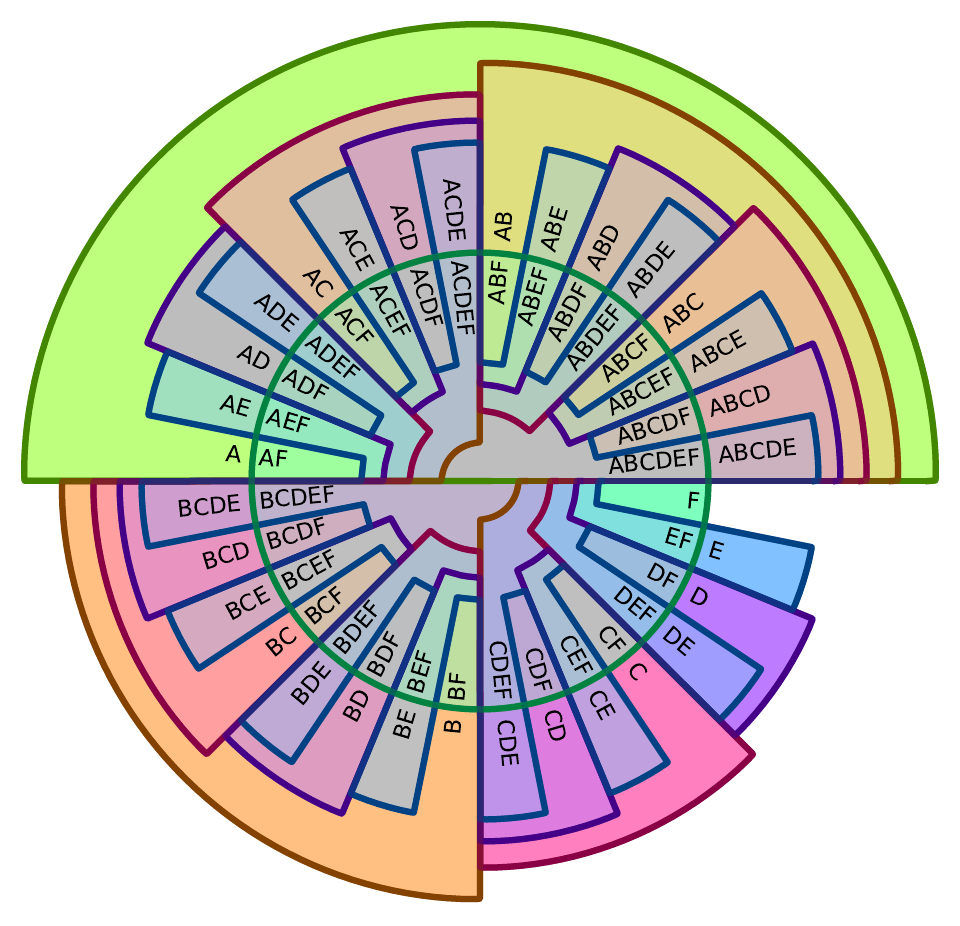}
        \subcaption{Sine-based.}
     \end{subfigure}
     \hfill
     \begin{subfigure}[b]{.45\textwidth}
        \centering
        \includegraphics[width=\textwidth]{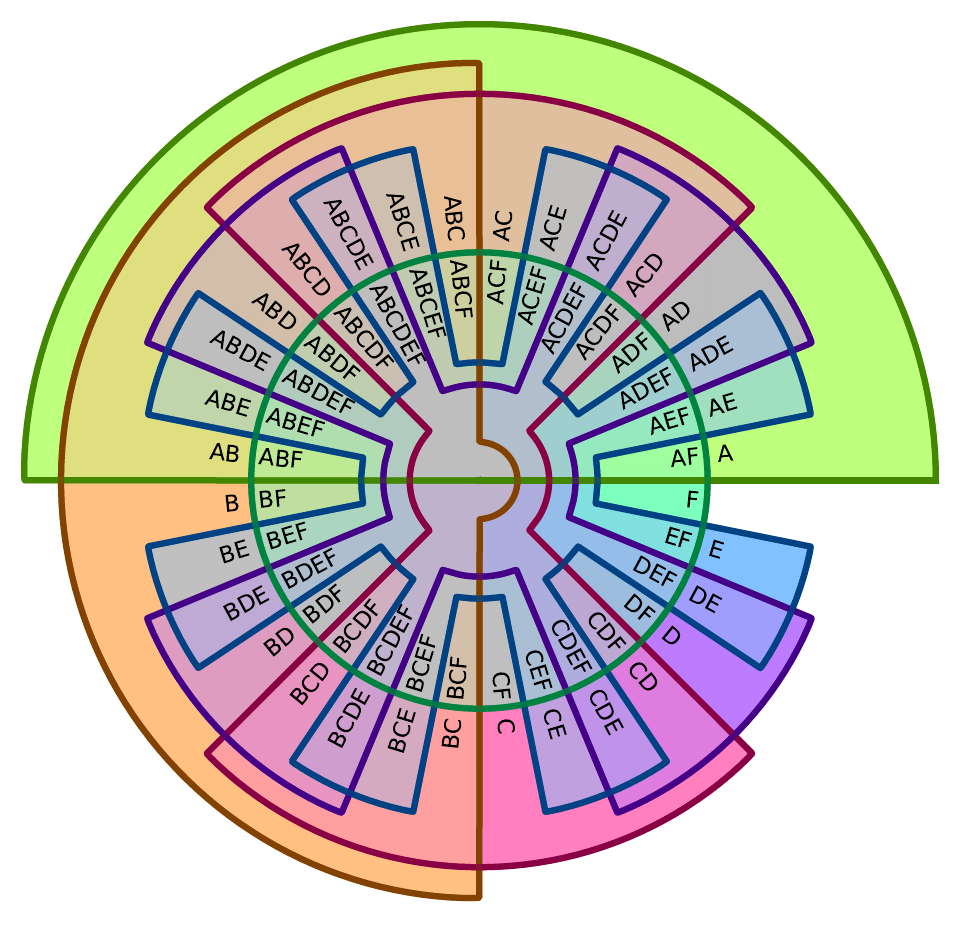}
        \subcaption{Cosine-based.}
     \end{subfigure}
     \caption{$n=6,\,p=1/1000,\,b=5/6$.}
     \label{fig:vennfan_extreme}
\end{figure}

One advantage of the \emph{VennFan} construction is the ability to adjust parameters to obtain diagrams with different visual styles.
For example, at very small $p$ we get a less rounded version of the diagram (Figure \ref{fig:vennfan_extreme}).
Arguably, this makes the sine-based version harder to parse, since it is not a simple diagram in the first place, and extremely small $p$ values exaggerate this.

\begin{figure}[H]
     \centering
     \begin{subfigure}[b]{.34\textwidth}
        \centering
        \includegraphics[width=\textwidth]{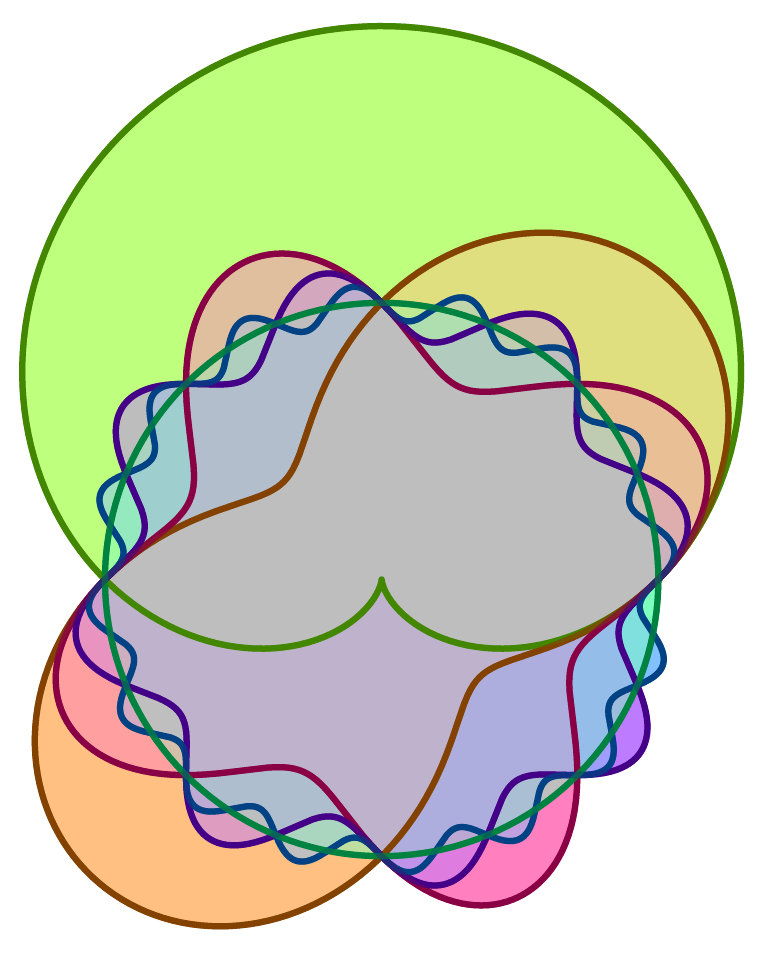}
        \subcaption{Sine-based.}
     \end{subfigure}
     \hfill
     \begin{subfigure}[b]{.5\textwidth}
        \centering
        \includegraphics[width=\textwidth]{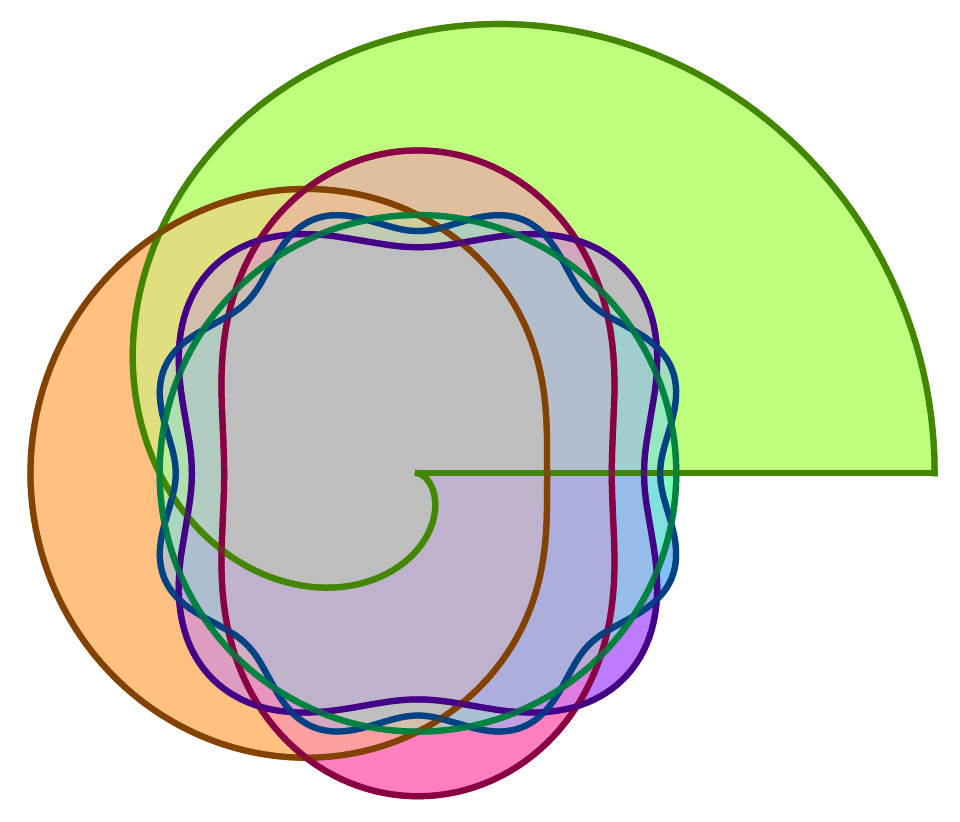}
        \subcaption{Cosine-based.}
     \end{subfigure}
     \caption{\emph{VennFan} constructions from unshaped sinusoids $n=6,\,p=1,\,b=1/2$.}
     \label{fig:vennfan_unmodified}
\end{figure}

Finally, to reiterate the importance of these modifications, in Figure \ref{fig:vennfan_unmodified} we show the projection of the classical trigonometric constructions.
They look beautiful, but they are not really suitable for representing data.

\subsection{Theoretical Remarks}
The radial projection step preserves the topology of the regions.
Intuitively, as long as the base curves $f_i$ form a valid diagram on the strip $[-\pi,\pi]\times[-1,1]$ (or $[2\pi,4\pi]\times[-1,1]$ for the cosine version) and we choose parameters such that $1+f_i(\vartheta)>0$ for all $i,\vartheta$, the curves $g_i(\vartheta)$ simply wrap the strip around the circle without tearing or gluing.
Consequently, the intersection pattern of the curves and the topological completeness of the Venn diagram is preserved by the projection.

The cosine family we start from is an explicit trigonometric realization of Edwards’ construction~\cite{edwards2004cogwheels, edwards1989new,weston2009symmetries}, which is known to produce simple $n$-Venn diagrams (no triple intersections) both on the sphere and, after projection, in the plane~\cite{ghouchan2022fullyreduciblesimplevenn,mamakanisimple}.
Hence the unmodified cosine construction underlying \emph{VennFan} is simple.
The cosine-based \emph{VennFan} with alternate scaling is often simple in practice, but this depends on the parameters (e.g.\ $p$ and the amplitude decay) and is not guaranteed a priori.
Moreover, the first boundary in the cosine construction ($\cos(x/2)$) does not complete a full period on $x\in[2\pi,4\pi]$, and its projection does not form a closed curve.
This can be remedied by connecting $g_0(2\pi)$ and $g_0(4\pi)$ with a straight line segment.

The sine-based \emph{VennFan} construction is not simple in general, and the curves may intersect more often than in the unmodified Smith construction, yielding an independent family rather than a strict Venn diagram.
However, for visualization purposes this is often not a practical issue: the boundaries are drawn with nonzero stroke width, and for aesthetically chosen parameters the extra tiny regions tend to be covered by the boundaries, so the rendered image is effectively a Venn diagram.

\section{Label placement}
    An important aspect of readable Venn diagrams is the placement of region labels.
    In \emph{VennFan}, we use two complementary strategies: radial labels placed near the unit circle, and labels placed by a geometric heuristic.
    The first option is to place labels around the projection circle in the angular order induced by the construction.
    As $x$ increases, the sign vector $(\operatorname{sgn}\,\sin(2^0 x),\dots,\operatorname{sgn}\,\sin(2^{n-1}x))$ changes by flipping exactly one coordinate at a time, yielding the binary-reflected Gray code order on $\{0,1\}^n$; this provides a principled circular ordering of the $2^n$ regions for radial labels~\cite{savage1997survey}.
    For larger $n$, a balanced approach would be to use radial placement for some regions and the visual-center heuristic for others.   

    We also consider defining and approximating an ``optimal'' text placement method inside each region.
    A straightforward heuristic would be to use the geometric center or \emph{centroid} of the region.
    For non-convex regions this can yield poor results: the centroid may lie very close to the boundary or even outside the region.
    A better heuristic is to place labels at points that approximate the \emph{visual center} of each region.
    Intuitively, the visual center is the point that ``looks most centered'' while remaining well separated from the boundary, and can be defined as the center of a largest inscribed disk in the region~\cite{garcia2007poles}.
    By standard compactness arguments, such a maximizer exists for every bounded region with nonempty interior.
    In practice we approximate it numerically (for example, using sampling or a distance transform on a rasterization of the diagram) and use this point as the anchor for the region’s text label.

    A limitation of the visual-center heuristic is that text labels are not point-like objects: they have width and height, and orientation matters.
    Simply placing the label at the visual center ignores rotation and can fail to maximize the readable font size.
    A more refined objective is to find an anchor point and a text rotation that maximize the font size for which the label still fits inside the region.
    
\begin{figure}[H]
         \centering
         \begin{subfigure}[b]{.46\textwidth}
            \centering
            \includegraphics[width=\textwidth]{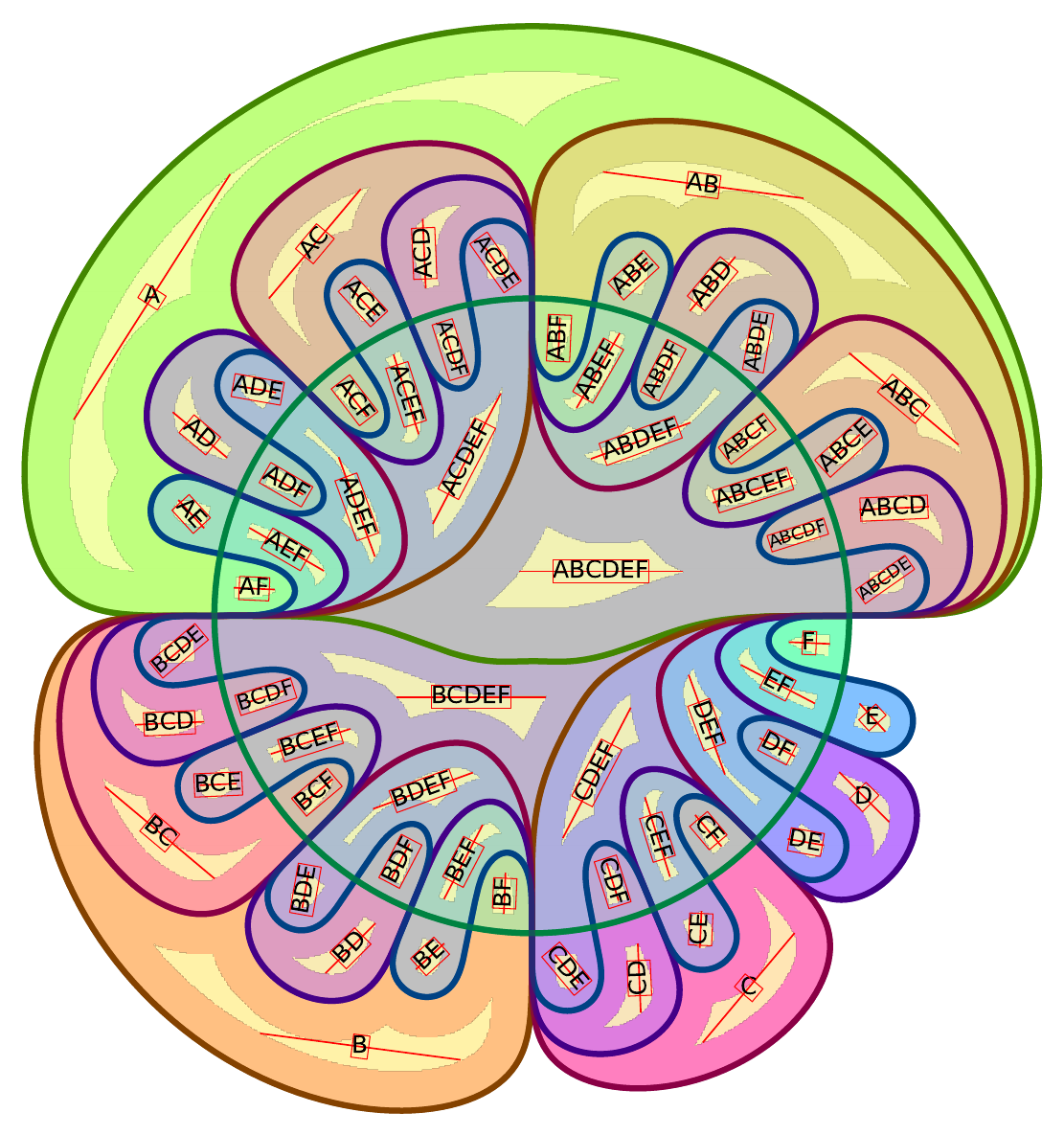}
            \subcaption{Sine-based.}
         \end{subfigure}
         \hfill
         \begin{subfigure}[b]{.52\textwidth}
            \centering
            \includegraphics[width=\textwidth]{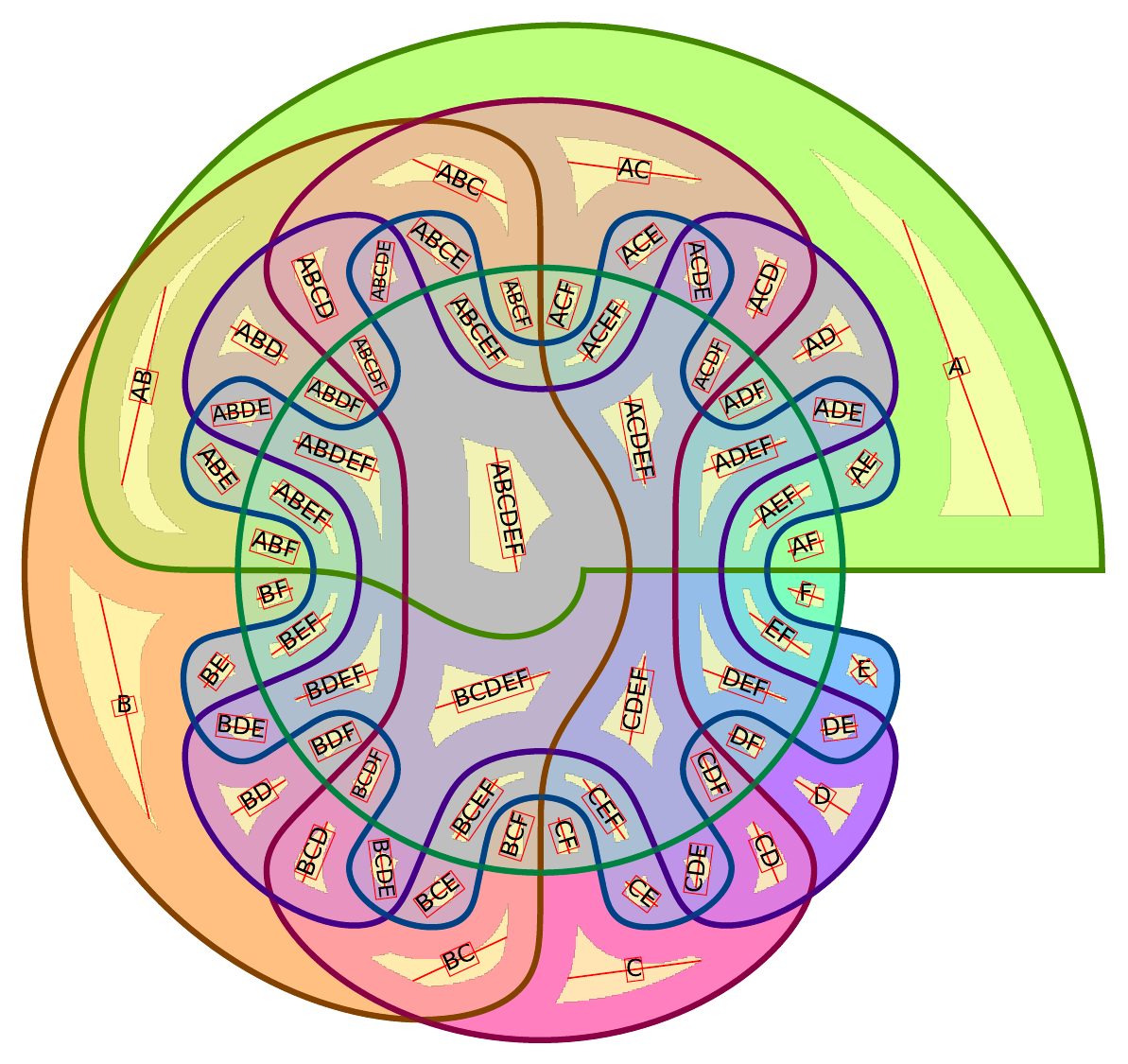}
            \subcaption{Cosine-based.}
         \end{subfigure}
         \caption{\emph{VennFan} label placement heuristic $n=6,\,p=1/5,\,\delta=1/4,\,\varepsilon=1/7$.}
         \label{fig:vennfan_N6_heuristic}
    \end{figure}

\begin{figure}[H]
         \centering
         \begin{subfigure}[b]{.64\textwidth}
            \centering
            \includegraphics[width=\textwidth]{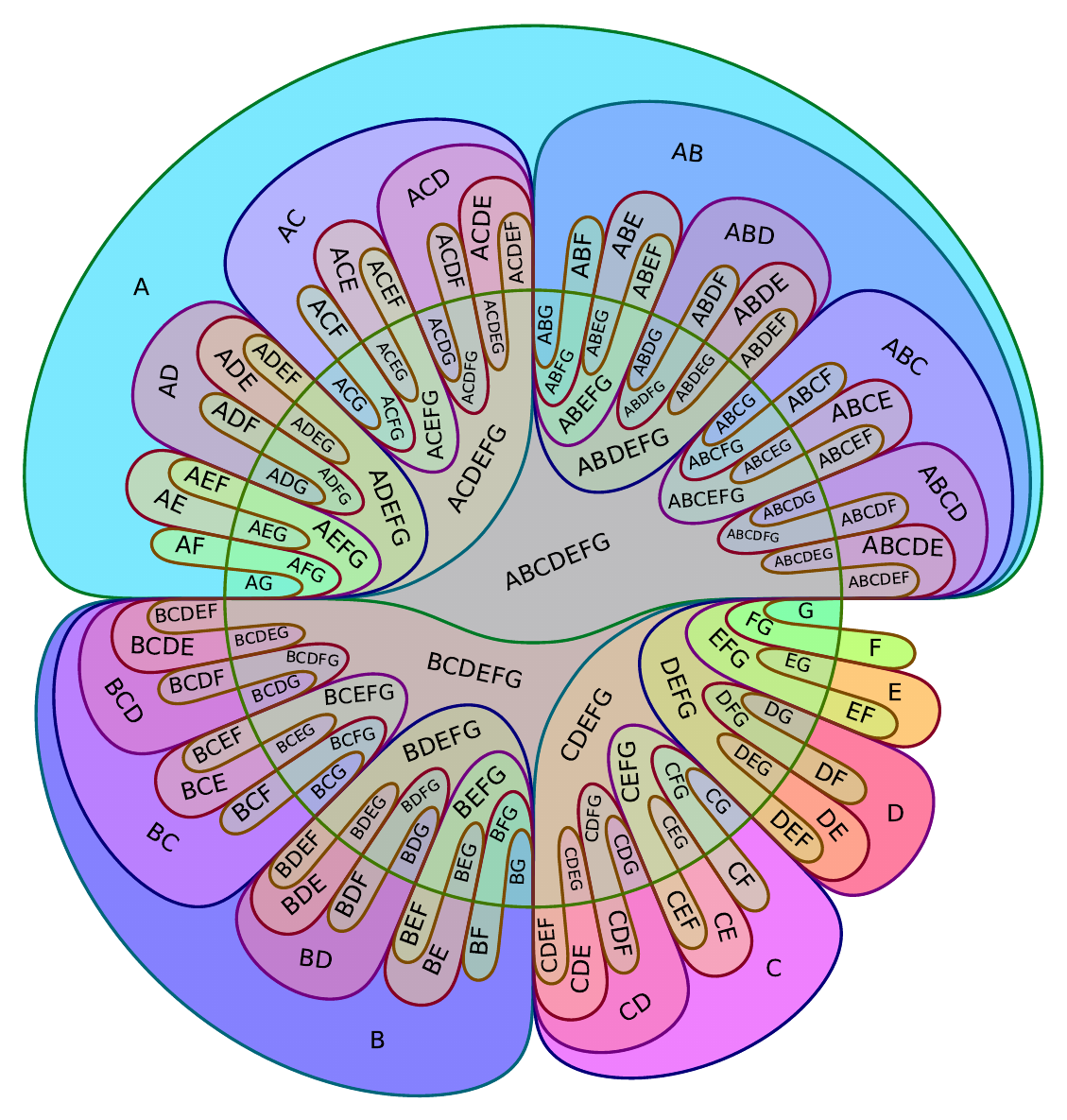}
         \end{subfigure}
         \caption{Sine-based \emph{VennFan} $n=7,\,p=1/7,\,\delta=1/4,\,\varepsilon=1/7$.}
         \label{fig:vennfan_sine_N7}
    \end{figure}

\begin{figure}[H]
         \centering
         \begin{subfigure}[b]{.64\textwidth}
            \centering
            \includegraphics[width=\textwidth]{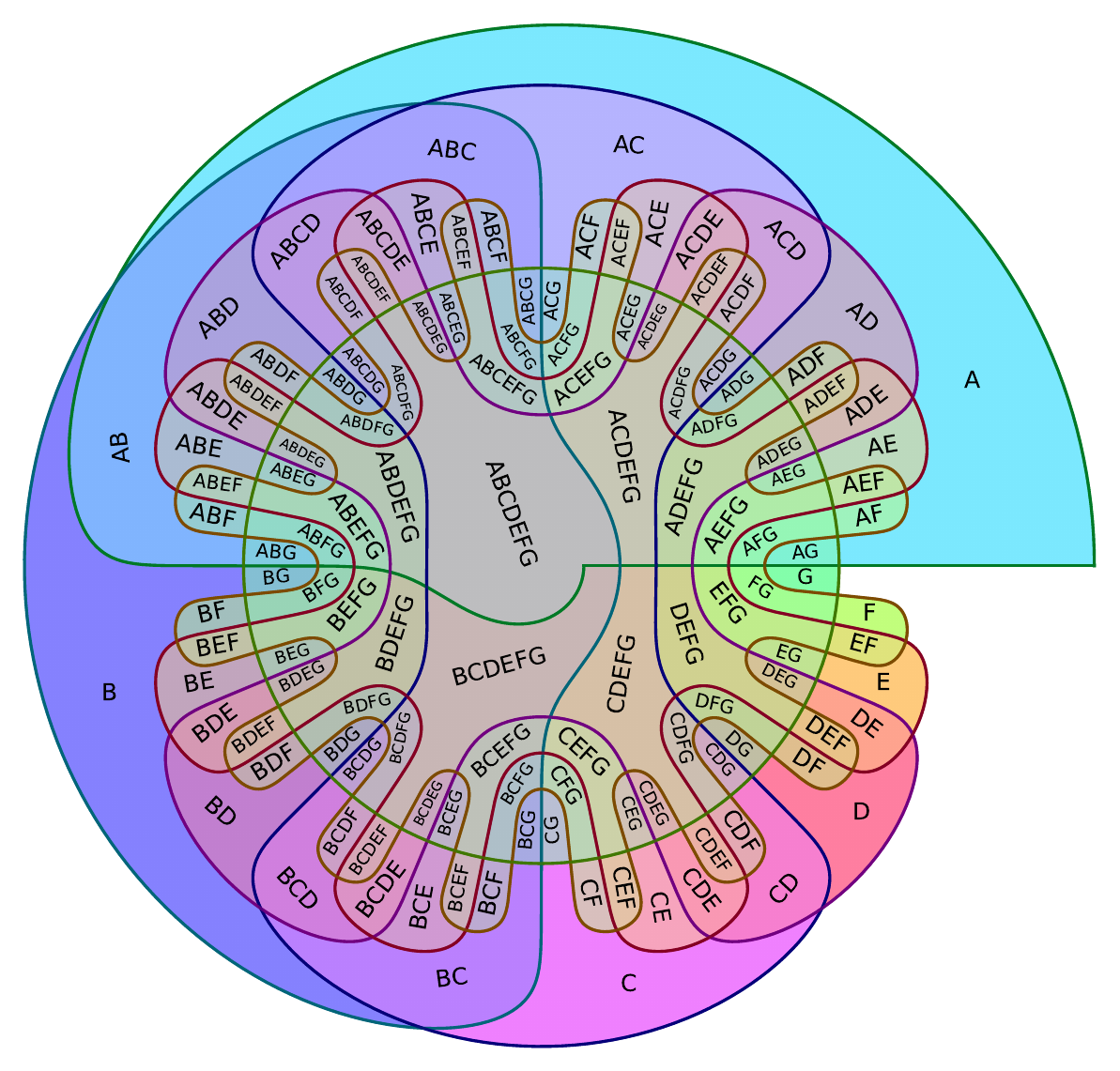}
         \end{subfigure}
         \caption{Cosine-based \emph{VennFan} $n=7,\,p=1/7,\,\delta=1/4,\,\varepsilon=1/7$.}
         \label{fig:vennfan_cosine_N7}
    \end{figure}

    We approximate this objective with the following heuristic.
    First, we erode the region until its area is reduced to a fixed fraction of its original area; this shrinks the region inward and creates a margin.
    On this eroded region we then look for the longest line segment that fits entirely inside it.
    In practice, we approximate this by taking the visual center of the eroded region and sampling line directions through it, keeping the longest segment that remains inside.
    Finally, we place the label at the midpoint of this segment and rotate the text to align with the segment direction.
    This produces labels that are visually centered in a more robust sense and can be rendered at larger font sizes without touching the boundary.
    Figure~\ref{fig:vennfan_N6_heuristic} shows the steps of this heuristic for $n=6$, and Figures~\ref{fig:vennfan_sine_N7} and \ref{fig:vennfan_cosine_N7} illustrate the resulting label placements for $n=7$.

\section{Conclusions}
    We presented \emph{VennFan}, a family of $n$-set Venn diagrams obtained from shaped trigonometric curves by polar coordinate projection to the disc. 
    We replaced the exponential amplitude decay of Smith’s construction by tunable decay schemes and obtained closed, fan-shaped curves. 
    In the cosine-based variant, we recover a planar diagram that is topologically equivalent to Edwards’ cogwheels, but with more direct geometric control.
    For suitable parameter choices, the resulting diagrams remain practically readable for $n$ up to about $8$.
    The accompanying Python implementation exposes the underlying parameters, allowing users to adjust the trade-off between visual styles to fit their specific use cases.




\section*{Declaration on Generative AI}
    During the preparation of this manuscript, OpenAI’s ChatGPT-5 was used to assist with grammar, paraphrasing, rewording, and targeted code review and augmentation.
    All AI-assisted outputs were subsequently reviewed, edited, and, where applicable, validated, and we assume full responsibility for the content of the publication.

\section*{Acknowledgements}
    Supported by the EKÖP-25 University Research Scholarship Program of the Ministry for Culture and Innovation from the source of the National Research, Development and Innovation Fund.
    
    Additional support by the Hungarian National Research, Development and Innovation Office within the framework of the Thematic Excellence Program 2021 -- National Research Sub programme: ``Artificial intelligence, large networks, data security: mathematical foundation and applications'' and the Artificial Intelligence National Laboratory Program (MILAB).
    
    We would like to express our gratitude to András Lukács for his valuable consultation.
    We would also like to thank GitHub for providing us academic access.
    
\newpage
\bibliographystyle{plain}
\bibliography{vennfan_references}

\newpage
\appendix
\section{Recreating Edwards' Cogwheel Diagrams}\label{appx:Edwards}
\begin{figure}[H]
         \centering
         \begin{subfigure}[b]{.48\textwidth}
            \centering
            \includegraphics[width=\textwidth]{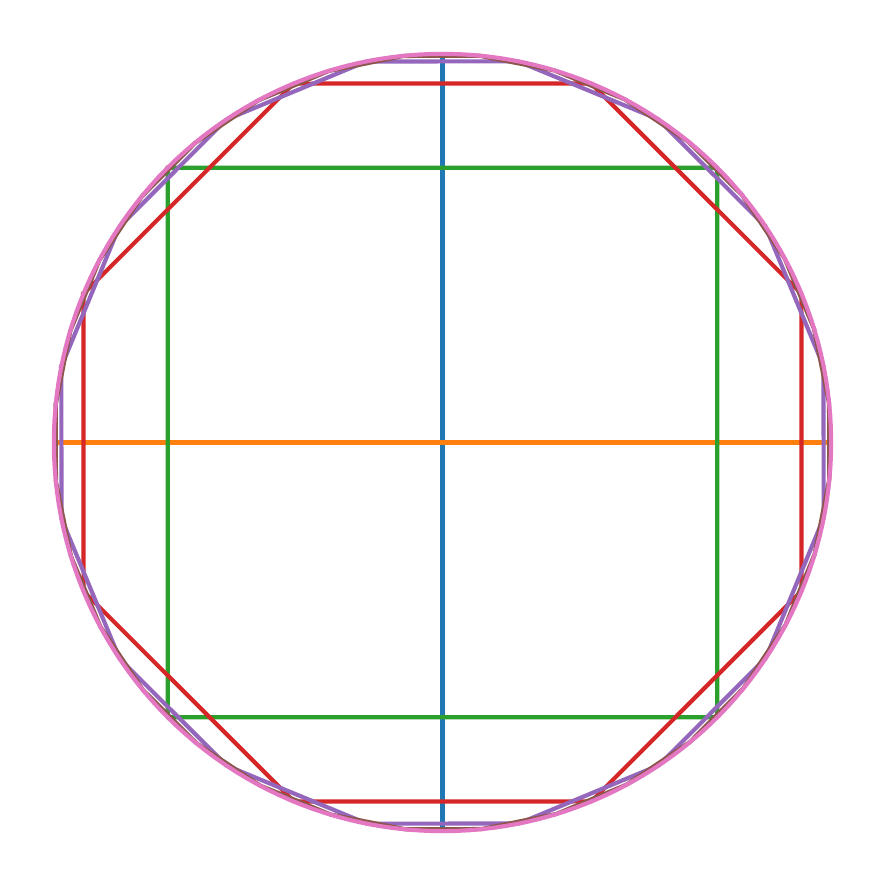}
            \subcaption{Equatorial projection of the curves (prism bases).}
            \label{fig:Edwards_sphere_equatorial}
         \end{subfigure}
         \hfill         
         \begin{subfigure}[b]{.46\textwidth}
            \centering
            \includegraphics[width=\textwidth]{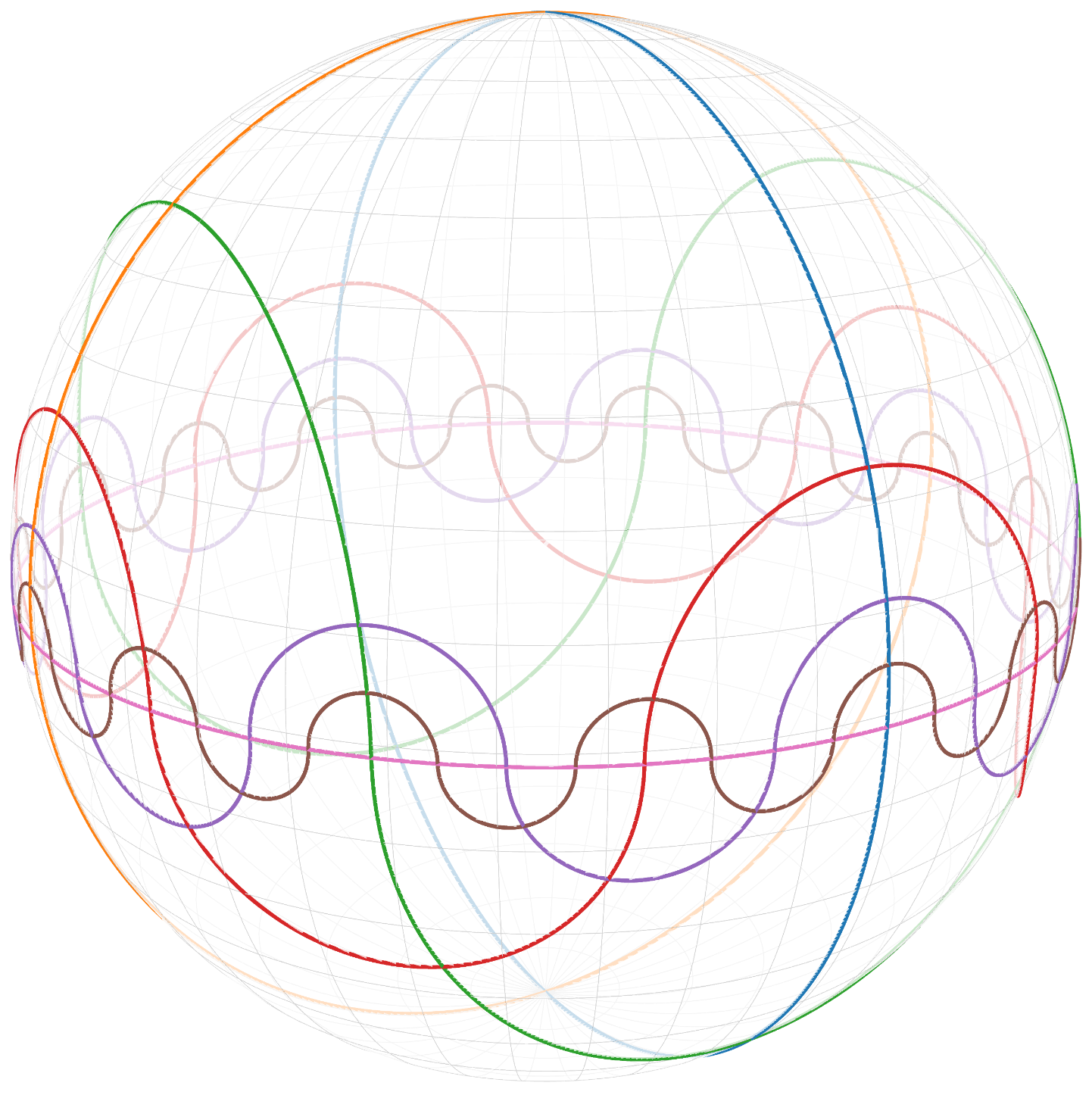}
            \subcaption{Edwards' Venn diagram on the sphere.}
            \label{fig:Edwards_sphere}
         \end{subfigure}
         \caption{Edwards' spherical Venn diagram and its equatorial projection ($n=7$).}
    \end{figure}

\begin{figure}[H]
         \centering
         \begin{subfigure}[b]{.48\textwidth}
            \centering
            \includegraphics[width=\textwidth]{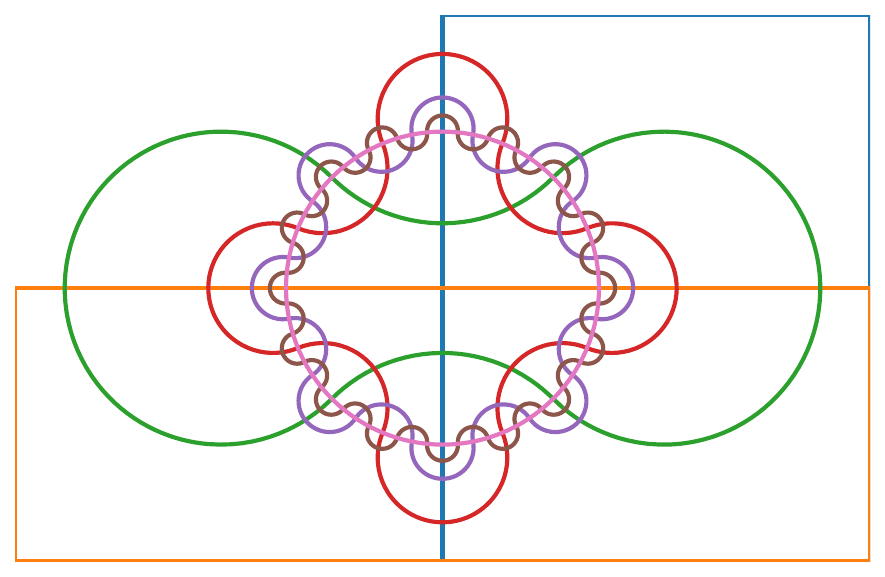}
            \subcaption{Projection onto a plane below the sphere.}
            \label{fig:Edwards_recreation_b}
         \end{subfigure}
         \hfill         
         \begin{subfigure}[b]{.48\textwidth}
            \centering
            \reflectbox{\includegraphics[width=\textwidth]{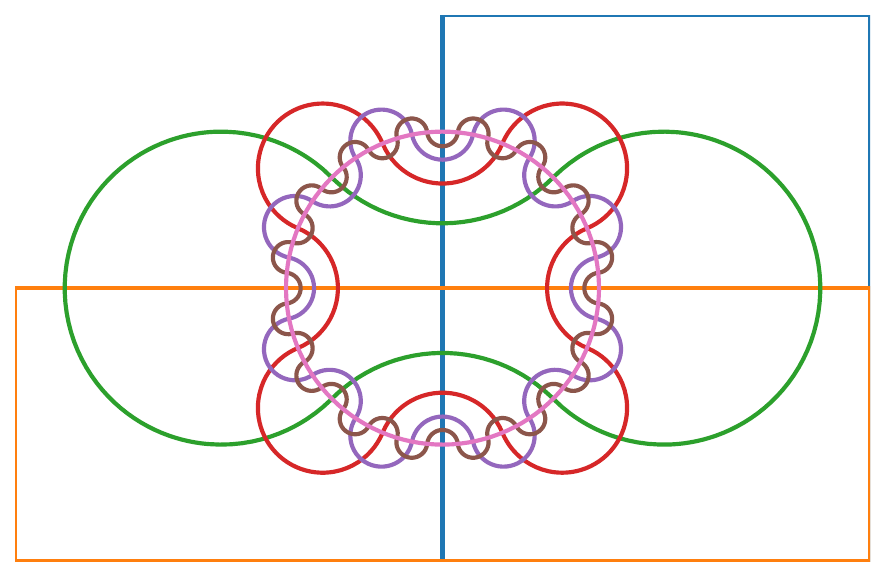}}
            \subcaption{Projection onto a plane above the sphere.}
            \label{fig:Edwards_recreation_a}
         \end{subfigure}
         \caption{Edwards' cogwheel Venn diagram obtained by stereographic projection ($n=7$).}
         \label{fig:Edwards_recreation}
    \end{figure}

    Edwards' Venn diagrams are based on a spherical construction.
    A convenient way to describe the construction is via prisms with regular $2^k$-gonal bases whose side faces intersect the sphere in circles.
    Figure~\ref{fig:Edwards_sphere_equatorial} shows these bases as they appear under orthogonal projection onto the equatorial disk.
    On the sphere there are three distinguished circles: the equator and two mutually orthogonal longitudes.
    In the equatorial projection these become one circle and two line segments (diameters).
    Each $2^k$-gonal prism intersects the sphere in $2^k$ circles; from these we select northern and southern semicircles in alternating order around the polygon.
    Together with the equator and the two longitudes, these arcs form a simple $n$-set Venn diagram on the sphere  (Figure~\ref{fig:Edwards_sphere}).
    Finally, stereographic projection of this spherical configuration yields the planar cogwheel diagrams of Edwards: Figure~\ref{fig:Edwards_recreation_b} shows the canonical variant, while Figure~\ref{fig:Edwards_recreation_a} shows the ``inside-out'' variant where the topological equivalence with \emph{VennFan} is more apparent.

\section{Additional \emph{VennFan} diagrams}

    \begin{figure}[H]
         \centering
         \begin{subfigure}[b]{.98\textwidth}
            \centering
            \includegraphics[width=\textwidth]{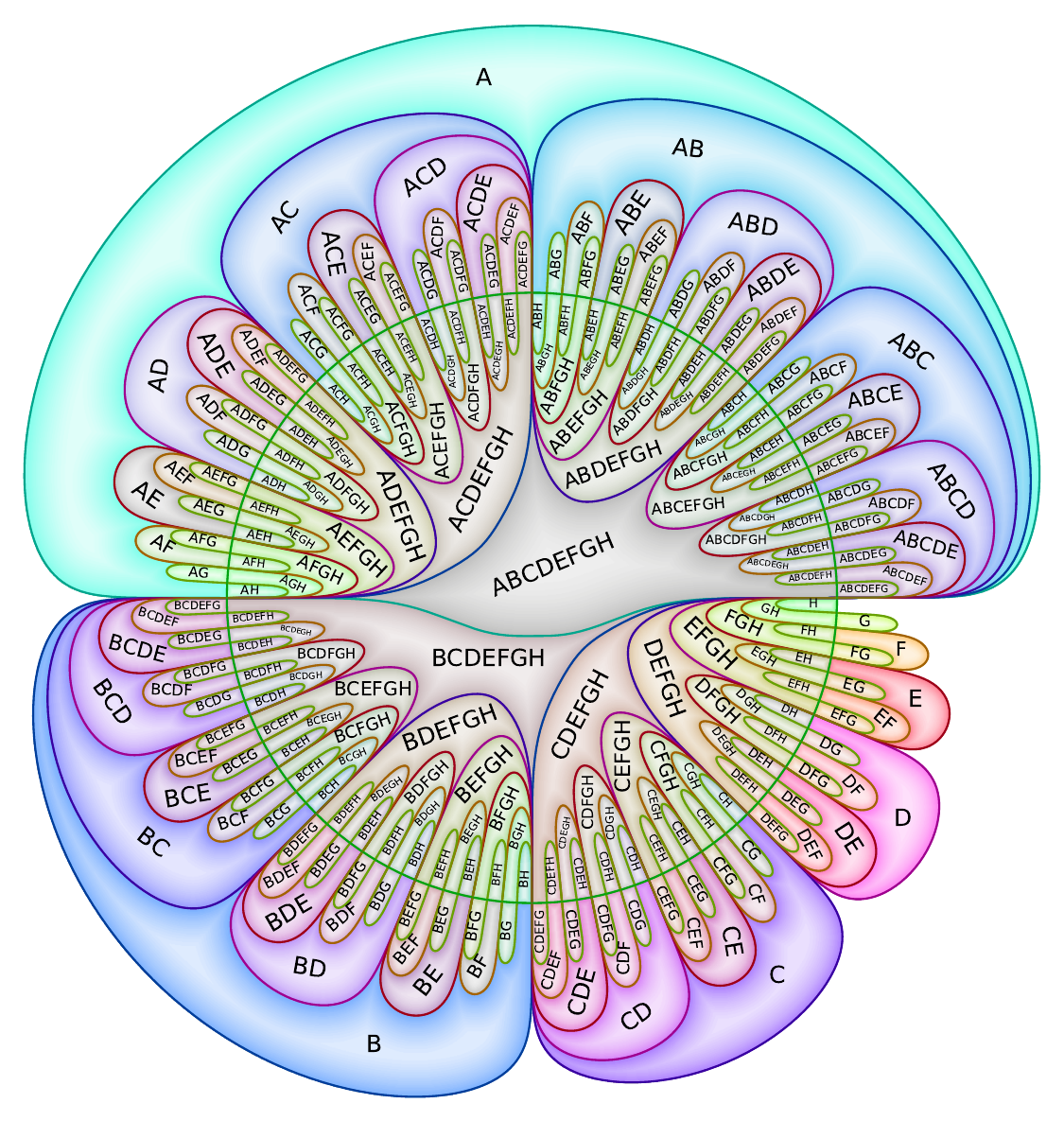}
         \end{subfigure}
         \caption{Sine-based \emph{VennFan} $n=8,\,p=1/7,\,\delta=1/5,\,\varepsilon=1/8$.}
         \label{fig:vennfan_sine_N8}
    \end{figure}
    
    \begin{figure}[H]
         \centering
         \begin{subfigure}[b]{.98\textwidth}
            \centering
            \includegraphics[width=\textwidth]{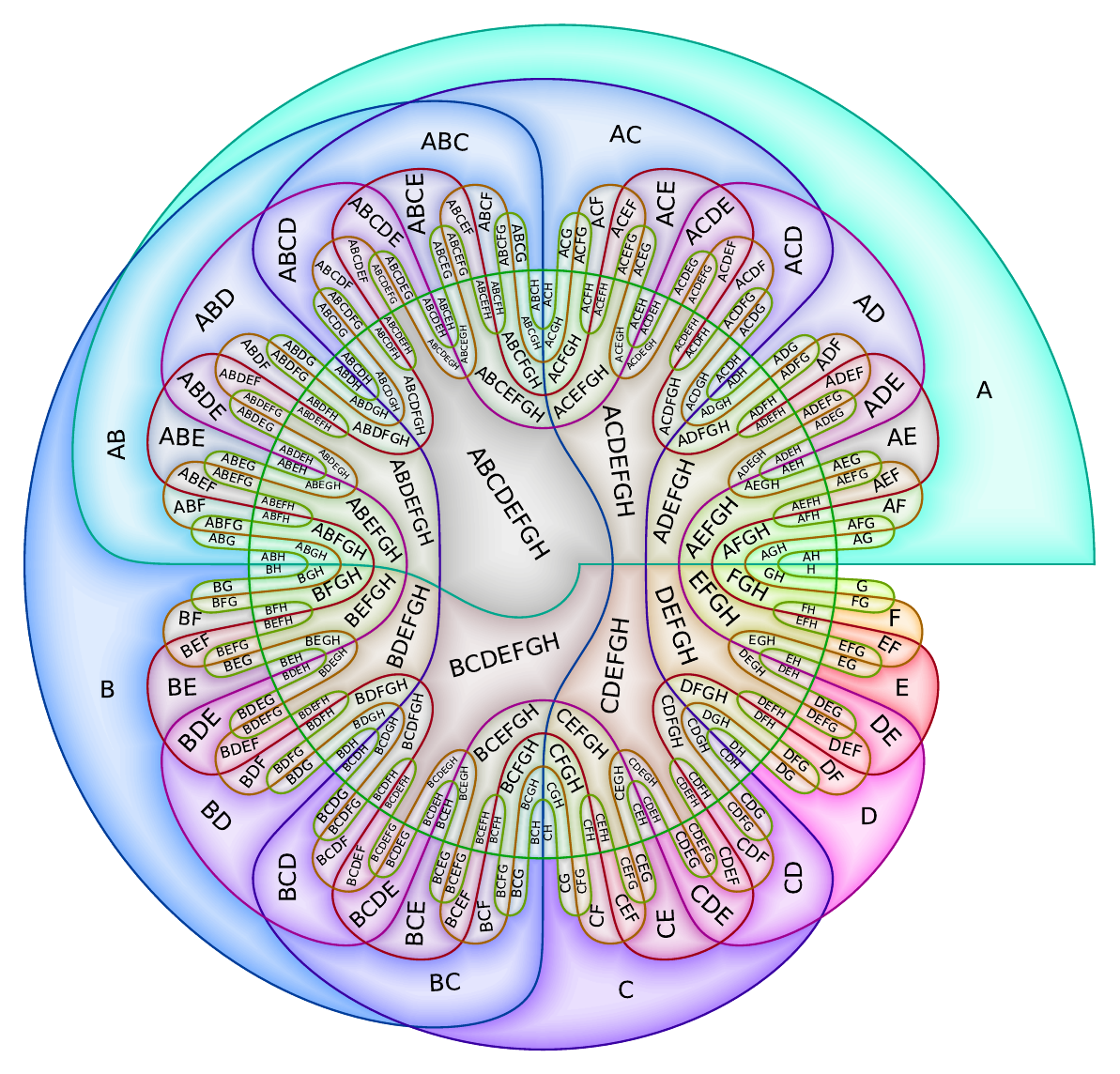}
         \end{subfigure}
         \caption{Cosine-based \emph{VennFan} $n=8,\,p=1/7,\,\delta=1/5,\,\varepsilon=1/8$.}
         \label{fig:vennfan_cosine_N8}
    \end{figure}

    \begin{figure}[H]
         \centering
         \begin{subfigure}[b]{.98\textwidth}
            \centering
            \includegraphics[width=\textwidth]{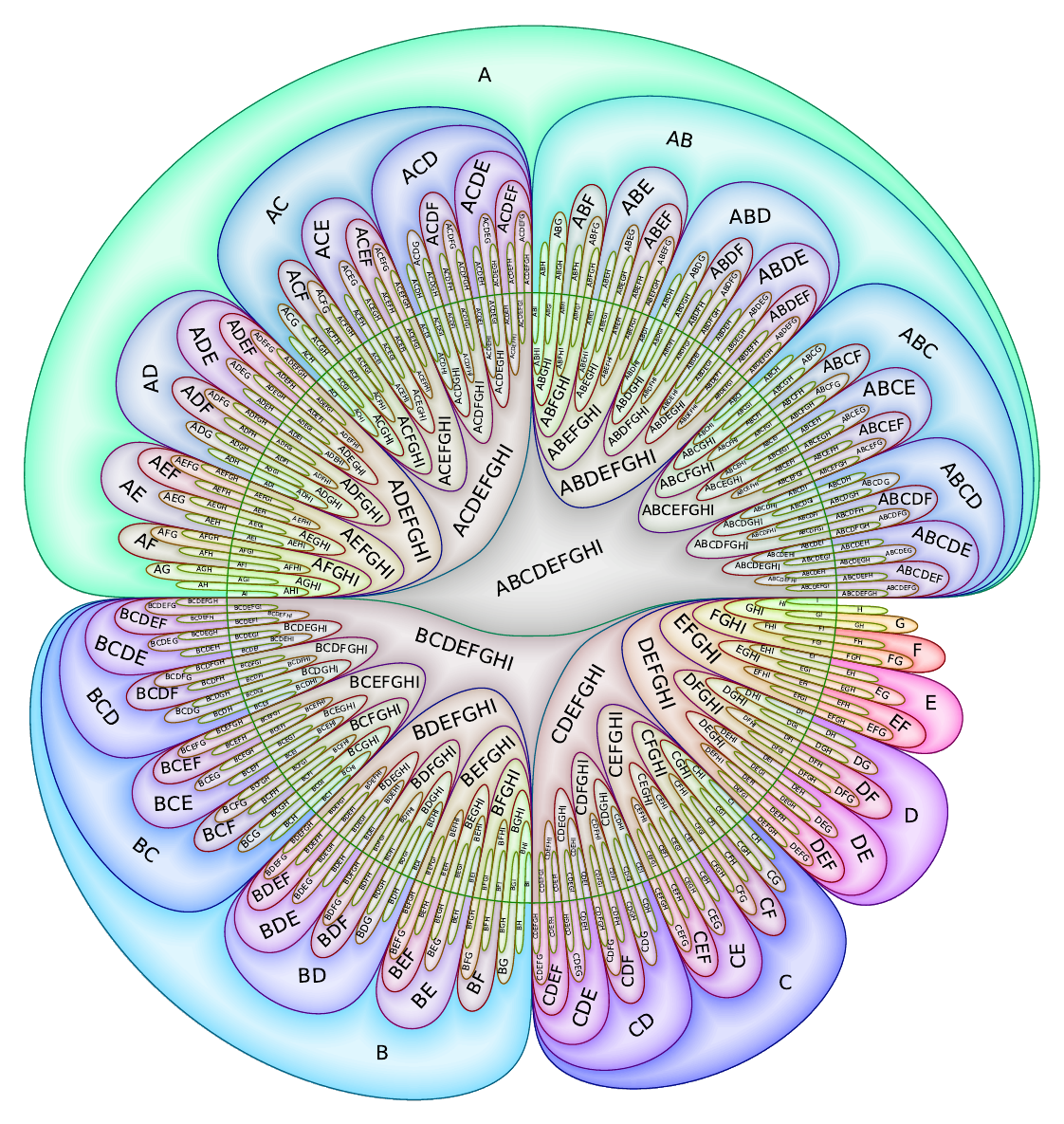}
         \end{subfigure}
         \caption{Sine-based \emph{VennFan} $n=9,\,p=1/7,\,\delta=1/6,\,\varepsilon=1/8$.}
         \label{fig:vennfan_sine_N9}
    \end{figure}
    
    \begin{figure}[H]
         \centering
         \begin{subfigure}[b]{.98\textwidth}
            \centering
            \includegraphics[width=\textwidth]{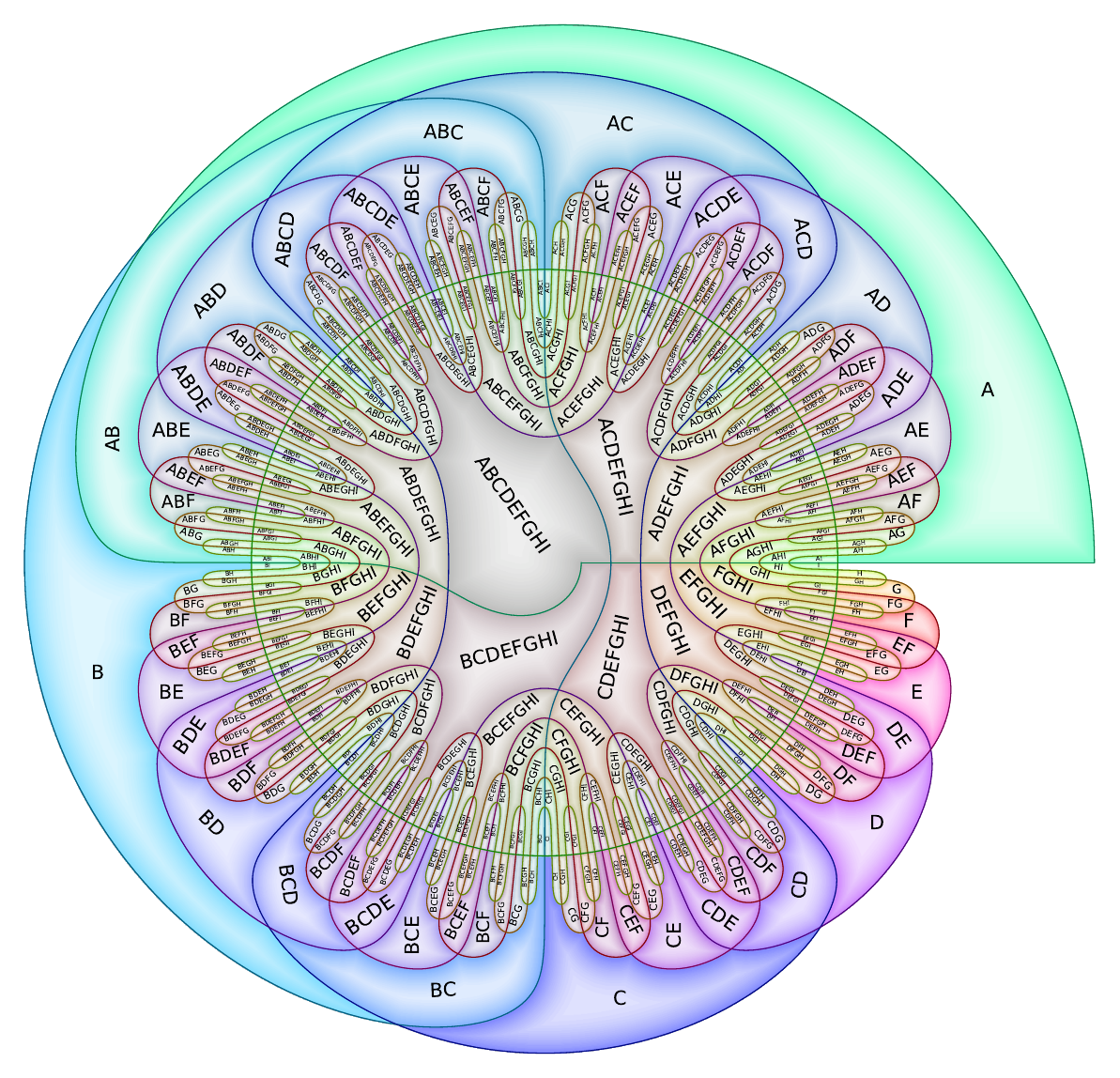}
         \end{subfigure}
         \caption{Cosine-based \emph{VennFan} $n=9,\,p=1/7,\,\delta=1/6,\,\varepsilon=1/8$.}
         \label{fig:vennfan_cosine_N9}
    \end{figure}
\end{document}